# Should a Normal Imputation Model Be Modified to Impute Skewed Variables?

Paul T. von Hippel

## Abstract *(169 words)*

Researchers often impute continuous variables under an assumption of normality, yet many incomplete variables are skewed. We find that imputing skewed continuous variables under a normal model can lead to bias; the bias is usually mild for popular estimands such as means, standard deviations, and linear regression coefficients, but the bias can be severe for more shape-dependent estimands such as percentiles or the coefficient of skewness. We test several methods for adapting a normal imputation model to accommodate skewness, including methods that transform, truncate, or censor (round) normally imputed values, as well as methods that impute values from a quadratic or truncated regression. None of these modifications reliably reduces the biases of the normal model, and some modifications can make the biases much worse. We conclude that, if one has to impute a skewed variable under a normal model, it is usually safest to do so without modifications—unless you are more interested in estimating percentiles and shape that in estimated means, variance, and regressions. In the conclusion, we briefly discuss promising developments in the area of continuous imputation models that do not assume normality.

*Key words*: missing data; missing values; incomplete data; regression; transformation; normalization; multiple imputation; imputation

Paul T. von Hippel is Assistant Professor, LBJ School of Public Affairs, University of Texas, 2315 Red River, Box *Y,* Austin, TX 78712, paulvonhippel.utaustin@gmail.com. I thank Ray Koopman for help with the calculations in Section 2.2. I also thank Pamela Paxton and Sheri Kunovich for sharing the data in Section 5.1

# 1 INTRODUCTION

Imputation is an increasingly popular method for handling data with missing values. When using imputation, analysts fill in missing values with random draws from an *imputation model*, and then fit the imputed data to an *analysis model*. In multiple imputation (MI), the process of imputation and analysis is repeated several times, and the results of the several analyses are combined (Rubin 1987; Allison 2002; Kenward and Carpenter 2007).

In an ideal world, the imputation model would perfectly represent the distribution of the data. But such perfect fidelity can be very difficult to achieve, and in practice it is often unnecessary. All that is necessary is that the imputation model preserve *those aspects of the distribution that are relevant to the analysis model*. For example, if the analyst plans only to estimates means, standard deviations, and the parameters of a linear regression model, then the imputation model needs only to preserve the means, variances, and covariances among the variables that will be analyzed.

When the goals of analysis are limited in this way, as they often are, a crude imputation model can yield usable results. For example, in some settings it can be acceptable to impute dummy variables, squared terms, and interactions as though they were normal when conditioned on other variables (Horton, Lipsitz, and Parzen 2003; Allison 2005; Bernaards, Belin, and Schafer 2007; von Hippel 2009). The resulting imputed values will look implausible if inspected closely, but that often has little effect on the analytic results. In fact, attempts to edit the imputed values to improve their plausibility can introduce bias by changing the variables' means, variance, and covariances (Horton et al. 2003; Allison 2005; Bernaards et al. 2007; von Hippel 2009).

The point of imputation is not that the imputed values should *look* like observed values. The point is that the imputed variable should *act* like the observed variable when used in analysis.

This paper considers methods for imputing a different class of non-normal variables—namely, variables with skew. Like other non-normal variables, skewed variables are often imputed as though they were conditionally normal. The word "conditionally" is important here; later we will encounter situations where a variable is skewed and yet the residuals are approximately normal when the skewed variable is conditioned on other variables. In that situation, a conditionally normal imputation model may be perfectly specified even though the imputed variable is skewed. In many other situations, though, imputing a skewed variable from a normal model entails some degree of misspecification.

In calculations and simulations, we find that conditionally normal imputation of a skewed variable can often produce acceptable estimates if the quantities being estimated are means, variances, and regressions. The estimates do have biases under some circumstances, but the biases are typically small. However, the biases grow much larger if we estimate quantities that depend more strongly on distributional shape—quantities such as percentiles or the coefficient of skewness.

To increase the skew of a normally imputed variable, popular references recommend modifications of the normal imputation model. Modifications include rounding (censoring) the imputed values, truncating the imputed values, imputing from a truncated regression model, or transforming the incomplete variable to better approximate normality. We evaluate all these methods, as well as a new method that adds quadratic terms to the imputation model.

In univariate data, we find that such modifications can work well if very carefully applied. But in bivariate and trivariate data, we find that even careful application of modified normal methods is problematic. At best, the modifications reduce the biases only a little; at worst, the modifications make the biases much worse. We conclude that, if you have to impute a skewed variable as though it were normal, it is safest to do so without modification.

In our conclusion we discuss the potential of non-normal imputation models that are currently in development. Early evaluations of these methods look promising, and we hope that they will soon become more widely available.

Our presentation proceeds in order of complexity, progressing from univariate to bivariate to multivariate data.

# 2 UNIVARIATE DATA

Imputation is rarely useful in univariate data since, if univariate values are missing at random, the observed values are a random sample from the population, and can be analyzed just as though the data were complete. Yet the simplicity of the univariate setting can help to clarify the advantages and disadvantages of different imputation methods.

Suppose we have a simple random sample of $n$ values from a non-normal variable $X$ with finite mean $\mu$ and finite variance $\sigma^2$. A randomly selected $n_{\text{mis}}$ of the $X$ values are missing, so that $n_{\text{obs}}=n-n_{\text{mis}}$ values are observed. We assume that values are *missing at random* (Rubin 1976; Heitjan and Basu 1996), which in the univariate setting means that the $n_{\text{obs}}$ observed $X$ values are a random sample from the $n$ sampled cases, and therefore a random sample from the population.

How well can normal imputation, with and without modifications, impute this non-normal variable?

## 2.1 Fully normal (FN) imputation

The first and simplest technique is Fully Normal (FN) imputation (Rubin and Schenker 1986), under which a nonnormal variable $X$ is imputed as though it were normal.

Since the observed values are a random sample from the population, we can obtain consistent estimates $\hat{\mu}_{obs}, \hat{\sigma}^2_{obs}$ from the observed values alone, and use those estimates to impute the missing values with random draws from a normal distribution $N(\hat{\mu}_{obs}, \hat{\sigma}^2_{obs})$. We then analyze the mix of imputed and observed values as though it were complete.

The consistency of the observed-value estimators $\hat{\mu}_{obs}$, $\hat{\sigma}^2_{obs}$ does not depend on $X$ being normal. The most obvious observed-value estimator are the mean $\bar{X}_{n_{obs}}$ and variance $s^2_{n_{obs}-1}$ of the observed values, which are minimum variance unbiased (MVU) estimators whether $X$ is normal or not (Halmos 1946). And the estimators that are commonly used in multiple imputation are obtained by supplementing $\bar{X}_{n_{obs}}$ and $s^2_{n_{obs}-1}$ with random variation to obtain Bayesian posterior draw (PD) estimators $\hat{\sigma}^2_{obs,PD}$ and $\hat{\mu}_{obs,PD}$—

$$\hat{\sigma}^2_{obs,PD} = s^2_{n_{obs}-1} \frac{n_{obs}-1}{U}, \text{where } U \sim \chi^2_{n_{obs}-1+\nu_{prior}}$$

$$\hat{\mu}_{obs,PD} \sim N\left(\bar{X}_{n_{obs}}, \frac{\hat{\sigma}^2_{obs,PD}}{n_{obs}}\right)$$

(1)

—which simulate random draws from the posterior density of $\mu, \sigma^2$ (Rubin and Schenker 1986). Note that $\nu_{prior}$ in equation (1) is the degrees of freedom in the Bayesian prior (von Hippel 2012).

The PD estimators, like the MVU estimators, are consistent.[1] However, the PD estimators are not particularly efficient and $\hat{\sigma}^2_{obs,PD}$ can be biased in small samples (von Hippel 2012).

We can spare ourselves further detail by assuming that we have a very large sample where $\hat{\mu}_{obs}$, $\hat{\sigma}^2_{obs}$ are practically indistinguishable from $\mu, \sigma^2$. In that case, we are effectively imputing missing values from $N(\mu, \sigma^2)$, and any biases we encounter are asymptotic—that is, any biases we encounter cannot be eliminated by increasing the sample size.

Figure 1 illustrates the large-sample properties of FN imputation in a setting where the observed data are skewed. Here the observed values come from a standard exponential density $X_{obs} \sim Exp(\mu)$, with mean $\mu = 1$ and variance $\mu^2 = 1$, while the imputed values $X_{imp} \sim N(\mu, \mu^2)$ come from a normal density with the same mean and variance as the observed exponential variable.

←Figure 1 **near here**→

Although the observed and imputed variables have the same mean and variance, they do not have the same distributional shape, and that implies that, in a mix of observed and imputed values, some quantities will be estimated with bias. For example, estimates of skewness will be biased toward zero since the imputed values have no skew. And the first 16 percentiles, at least, will be negatively biased since 16% of the imputed distribution is negative, while the observed distribution is strictly nonnegative. The lower panel of Figure 1 examines all the percentiles by comparing the cumulative distribution function (CDF) of the observed and imputed variables. Percentiles 24-89.5 are positively biased since the imputed CDF is right of the observed CDF, but all the other percentiles are negatively biased since the imputed CDF is left of the observed CDF. In some other distributions, FN imputation of non-normal variables causes more bias in the extreme percentiles than in percentiles near the median (Demirtas, Freels, and Yucel 2008), but in this skewed example the 50[th] percentile has substantial bias while the 90[th] percentile has almost none.

---

[1] To see this, note that as $n \to \infty$, $\bar{X}_{n_{obs}}$ and $s^2_{n_{obs}-1}$ converge to $\mu$ and $\sigma^2$, while $(n_{obs}-1)/U$ and $\hat{\sigma}^2_{obs,PD}/n_{obs}$ converge to 1 and 0. Therefore $\hat{\mu}_{obs,PD}$ and $\hat{\sigma}^2_{obs,PD}$ converge to $\mu$ and $\sigma^2$—i.e., the PD estimators are consistent.

In sum, when non-normal data are imputed under a normal model, estimates of the mean and variance will be consistent, but there can be considerable bias in estimating the shape and percentiles. Simulations confirm that FN imputation yields consistent estimates for $\mu$ and $\sigma^2$ under a variety of non-normal distributions (Rubin and Schenker 1986; He and Raghunathan 2006; Demirtas and Hedeker 2008), but yields biased estimates of many percentiles (Demirtas and Hedeker 2008; He and Raghunathan 2006; Demirtas et al. 2008).

## 2.2   Imputing within bounds: censoring and truncation

What can make the observed and imputed distributions more similar? A popular approach is to bound normally imputed values within a plausible range. For example, in imputing a positively skewed variable like body weight, we might require the imputed values to be larger than the smallest observed body weight, or not so small as to be "biologically implausible" (Centers for Disease Control and Prevention 2011). Options for bounding imputed variables are available in most popular imputation software, including IVEware, the MI procedure in SAS 9.2, the *mi impute* command in Stata 12, and the Missing Values Analysis package in SPSS 16.0.

There are two general approaches to bounding: censoring and truncation. In Figure 1, for example, where we imputed a standard exponential variable as though it were normal, we could have used either censoring or truncation to require the normally imputed values to be non-negative. To do this, we would first generate normal imputations $X_{imp} \sim N(\mu_{pre}, \sigma^2_{pre})$ where $\mu_{pre}, \sigma^2_{pre}$ are the mean and variance of the imputed variable before bounds have been imposed. We would then round negative imputed values up to zero (censoring), or re-impute negative values repeatedly until a positive value occurs (truncation). Truncation results in a distribution of imputed values that is strictly positive, while censoring results in a distribution that is strictly positive except for a point mass at $X_{imp,censor} = 0$.

Bounding imputed values changes the mean and variance of the imputed variable. After bounding, the mean of the imputed variable is higher than $\mu_{pre}$ because negative values have been shifted upward, and the variance of the imputed variable is smaller than $\sigma^2_{pre}$ because out-of-bounds values have been shifted inward, away from the boundary. More specifically, if an imputed normal variable $X_{imp} \sim N(\mu_{pre}, \sigma^2_{pre})$ is *truncated* to the left of $c$, the truncated variable has mean and variance

$$E(X_{imp,trunc}) = \mu_{pre} + \lambda_c \sigma_{pre}$$
$$V(X_{imp,trunc}) = (1 - \delta_c)\sigma^2_{pre}$$

(2)

And if the same imputed normal variable is *censored* to the left of $c$, the mean and variance are

$$E(X_{imp,censor}) = \pi_c c + (1 - \pi_c)E(X_{imp,trunc})$$
$$V(X_{imp,censor}) = (1 - \pi_c)(V(X_{imp,trunc}) + (\lambda_c - Z_c)^2 \pi_c \sigma^2_{pre})$$

(3)

Here $Z_c = (c - \mu_{pre})/\sigma_{pre}$ is the standardized boundary $c$; $\phi(Z_c)$ and $\pi_c = \Phi(Z_c)$ are the standard normal PDF and CDF evaluated at $Z_c$; $\lambda_c = \phi(Z_c)/(1 - \Phi(Z_c))$) is the Mills ratio; and $\delta_c = \lambda_c(\lambda_c - Z_c)$ (e.g., Jöreskog 2002; Greene 1999, p. 907; Johnson, Kotz, and Balakrishnan 1994, section 10.3).

Bias can occur if we fail to anticipate the effect of bounding on the mean and variance of the imputed variable. To return to our running example, if the observed variable is standard exponential $X_{obs} \sim Exp(\mu)$, with mean $\mu = 1$ and variance $\mu^2 = 1$, then we might naively impute a normal variable with the same mean and variance $\mu_{pre} = \sigma^2_{pre} = 1$. But if we censor this normal variable to the left of $c = 0$, the mean of the censored variable will be $E(X_{imp,censor}) \approx 1.08$ and the variance will be $V(X_{imp,censor}) \approx 0.75$—that is, the mean will have a bias of 8% and the variance will have a bias of –25%. The biases will be even larger if we *truncate* the normal imputed variable to the left of $c = 0$; then the mean of the truncated variable will be $E(X_{imp,trunc}) \approx 1.29$, which is biased by 29%, and the variance will be $V(X_{imp,trunc}) \approx 0.63$, which is biased by –37%.

These biases can be avoided, or at least reduced, if we anticipate the effect of censoring on the mean and variance of the imputed variable. By inverting equations (2) and (3) we can choose $\mu_{pre}, \sigma^2_{pre}$ so that, *after* bounds are imposed, the mean and variance of the bound imputations will be the same as those of $X_{obs}$, or at least close. Let's return to our example where the observed variable $X_{obs} \sim Exp(\mu)$ is standard exponential with mean $\mu = 1$ and variance $\mu^2 = 1$. If we want to get the same mean and variance from a censored normal variable that is bound to the left of $c=0$, we can do so by choosing $\mu_{pre} \approx .785$ and $\sigma_{pre} \approx 1.29$; then, after truncation, we will have a censored variable with a mean and variance of $E(X_{imp,censor}) = V(X_{imp,censor}) = 1$.

Similarly, if we want to get a similar mean and variance from a truncated normal variable that is bound to the left of $c=0$, we can get close by choosing $\mu_{pre} = -6.28$ and $\sigma_{pre} = 3.02$. Then after truncation we will have a truncated variable with a mean of $E(X_{imp,trunc}) = 1.1$ and a variance of $V(X_{imp,trunc}) = 1$.

Notice that, although the censored normal imputations have the same mean as the observed variable, the truncated normal imputations do not. In fact, with $c=0$, it is impossible to get the mean and variance of a censored normal variable to equal $E(X_{imp,trunc}) = V(X_{imp,trunc}) = 1$ exactly. As we approach $E(X_{imp,trunc}) = V(X_{imp,trunc}) = 1$, $\mu_{pre}/\sigma_{pre}$ approaches $-\infty$, so in the area near $E(X_{imp,trunc}) = V(X_{imp,trunc}) = 1$ it is very hard to estimate the parameters $\mu_{pre}, \sigma_{pre}$ with any precision.

We have just seen that, even in a simple univariate setting, a truncated normal model can have trouble matching the moments of a variable that is not in fact truncated normal. When the truncated normal model is used for this purpose, the parameter estimates can be sensitive or even infinite. Later we will encounter similar problems when we use the truncated model in a multivariate setting.

Figure 2 illustrates our attempts to match the distribution of a standard exponential variable with a truncated or censored normal variable. The figure illustrates a naïve approach, where the mean and variance of the imputed variable are unbiased before censoring or truncation, but biased after. The figure also illustrates less-biased approaches, where the effect of truncation or censoring are anticipated before the imputations are drawn. Notice that, if the effects of truncation are anticipated, the truncated normal variable achieves a close match to the exponential distribution (lower left)—notwithstanding some bias in the mean.

## 2.3   Transformation

Instead of bounding imputed *X* values, some experts recommend *transforming* an incomplete skewed *X* variable to better approximate normality (Schafer and Olsen 1998, p. 550; Schafer and Graham 2002, p. 167; Allison 2002, p. 39; Raghunathan et al. 2001, pp. 82-83). Following this advice, we would subject the observed values $X_{obs}$ to a skew-reducing transformation such as a log or root, which yields a transformed variable $X_{obs,t} = t(X_{obs})$ with a mean and variance of $\mu_t, \sigma_t^2$. We then apply FN imputation to the transformed variable to obtain normal imputed values $X_{imp,t} \sim N(\mu_t, \sigma_t^2)$. Finally, we invert the transformation to recover the original observed values $X_{obs} = t^{-1}(X_{obs,t})$ and put the imputed values $X_{imp} = t^{-1}(X_{imp,t})$ on a comparable scale. SAS automates this process in the MI procedure, where the TRANSFORM command offers a choice among a variety of nonlinear transformations including the log, power, Box-Cox (1964), logit, and exponential transformations (SAS Institute 2001).

In the univariate setting, transformation works well if the transformed variable is in fact normal or close to it. For example, if we apply a log transformation to a lognormal variable, the transformed variable is exactly normal so the transformation method is optimal.

But transformation can yield substantial bias if the transformed variable is not close to normal. To return to our earlier example, if we sample an exponential variable $X_{obs} \sim Exp(\mu)$ and transform it with a log, the logged variable $X_{obs,t} = ln(X_{obs})$ is far from normal (Hawkins and Wixley 1986). If we impute a normal variable $X_{imp,t} \sim N(\mu_t, \sigma_t^2)$ on the log scale, inverting the transformation will yield an imputed $X_{imp} = exp(X_{imp,t})$ whose mean and variance are far from the mean and variance of the original exponential variable (He and Raghunathan 2006).

The challenge of transformation is that skew-reducing transformations are nonlinear, and nonlinear transformations do complicated things to the mean and variance. Although on the transformed scale the observed variable $X_{obs,t}$ and the imputed variable $X_{imp,t}$ have approximately the same mean and variance $\mu_t, \sigma_t^2$, after the transformation is inverted there is no guarantee that $X_{obs}$ and $X_{imp}$, unless they have the same distribution, will also have the same mean and variance $\mu, \sigma^2$.

The transformation method can yield better results if the transformation is carefully chosen. In imputing an exponential variable $X_{obs} \sim Exp(\mu)$, for example, it is not hard to see why the logarithm is a poor choice. The log of an exponential variable is undefined at $X_{obs} = 0$, and although a value of exactly zero will never occur in a sample, values close to zero occur frequently, and the logs of these near-zero values will be large negative numbers. So the log of a standard exponential sample has a long left tail and is far from normal.[2]

Figure 3 illustrates two transformations that do a better job of normalizing an exponential variable $X_{obs}$: the square root transformation $\sqrt{X_{obs}}$ and the fourth root transformation $\sqrt[4]{X_{obs}}$. Unlike the log, both the square root and the fourth root are defined at $X_{obs} = 0$. The square root transformation is common in data analysis, while the fourth root is less common but comes about as close to normalizing the exponential distribution as any transformation can (Hawkins and Wixley 1986).

←Figure 3 **near here**→

Both the square-root and fourth-root transformations yield an imputed variable whose mean is unbiased, or nearly unbiased, when compared to the mean of an exponential variable. The variance of the imputed variable is biased, though less biased for the fourth root than for the square root.

More specifically[3], if we subject an observed standard exponential variable $X_{obs} \sim Exp(1)$ to a square root transformation, the result is a transformed variable $\sqrt{X_{obs}} = X_{obs,t} \sim Ray(1/\sqrt{2})$ that has a Rayleigh distribution with mean $\mu_t = \sqrt{\pi}/2$ and variance $\sigma_t^2 = 1 - \pi/4$. On the transformed scale, we can generate a normal imputed variable $X_{imp,t}$ with the same mean $\mu_t$ and variance $\sigma_t^2$. Finally, we invert the transformation, squaring all the values to recover the original observed variable $X_{obs} = X_{obs,t}^2 \sim Exp(1)$ and produce an imputed variable $X_{imp} = X_{imp,t}^2$ that, as the square of a normal variable, has a scaled noncentral chi-square distribution, with a mean of $E(X_{imp}) = 1$ and a variance of $V(X_{imp}) = 2 - \pi^2/8 \approx .77$. The mean of $X_{imp}$ is unbiased, but the variance of $X_{imp}$, when compared to the variance $V(X_{obs}) = 1$ of the observed variable, is negatively biased by 23%.

We get better results with a fourth-root transformation, although there is still a small bias. If the observed variable $X_{obs} \sim Exp(1)$ has a standard exponential distribution, then the fourth-root transformed variable $X_{obs,t} = \sqrt[4]{X_{obs}}$ is very close to normal with a mean of $\mu_t \approx .91$ and a variance of $\sigma_t^2 \approx .065$. If we impute a normal variable on the transformed scale $X_{imp,t} \sim N(\mu_t, \sigma_t^2)$ with the same mean and variance as $X_{obs,t}$, then when we invert the transformation the imputed variable $X_{imp} = X_{imp,t}^4$ has a mean of 1.006 and a variance of 1.135. The mean of $X_{imp}$ has less than 1% bias, while the variance of $X_{imp}$, is positively biased by 13.5%. The shape of the imputed distribution closely matches the observed variable except in the range very close to $X$=0 (Figure 3).

In short, in univariate data the use of normalizing transformation to impute skewed variables can sometimes yield good results if the transformation is carefully chosen. However, we still cannot recommend transformation since it has disadvantages in bivariate settings, which we will shortly discuss.

---

[2] In simulations, we found that the skew of a logged exponential sample was about– 1.14, but the true skew is undefined since $ln(0)$ is undefined.
[3] The following calculations were assisted by Mathematica 8.

# 3   BIVARIATE DATA

Bivariate data present new challenges for the imputation of skewed variables. In bivariate data, the imputation model should preserve not just the marginal distribution of the skewed variable; the imputation must also preserve aspects of the relationship between the skewed variable and other variables. This can be difficult.

To set the stage, suppose that we have, again, a standard exponential variable $X \sim Exp(1)$, but now suppose that we also have a second variable $Y$ that fits a linear regression on $X$ with normal residuals:

$$Y = \alpha_{Y|X} + \beta_{Y|X}X + e_{Y|X}, \text{ where } e_{Y|X} \sim N(0, \sigma^2_{Y|X}) \qquad (4)$$

Figure 4 displays $n$=500 simulated observations on $(X,Y)$ with $\alpha_{Y|X} = \beta_{Y|X} = 1$ and $R^2_{Y|X} = .7$. The figure shows a scatterplot for $Y$ regressed on $X$ with an OLS line (left) and a scatterplot for $X$ regressed on $Y$ (right) with an OLS line and an OLS quadratic.

←Figure 4 **near here**→

In evaluating imputation methods, it will be important to realize that the regression of $Y$ on $X$ satisfies the OLS assumptions, but the regression of $X$ on $Y$ does not. In particular, the regression of $X$ on $Y$ violates the OLS assumption that the conditional expectation of the residuals $E(e_{X|Y}|Y)$ is zero for all $Y$. To the contrary, $E(e_{X|Y}|Y) > 0$ when $Y$ is small and when $Y$ is large; in the scatterplot this is evident from the fact that all the $X$ values are above the OLS line when $Y$ is small or large. The residuals are non-normal as well, but this is less important since residual normality is not required for OLS estimates to be consistent (Wooldridge 2001).

In fact, the regression of $X$ on $Y$ is not even linear. When $X$ is non-normal, a linear regression of $Y$ on $X$ does not necessarily imply a linear regression of $X$ on $Y$. As Figure 4 show, the fit of the regression of $X$ on $Y$ can be improved, though not perfected, by adding a quadratic term $Y^2$ to the regression. With the quadratic term added, the regression comes closer to satisfying the OLS assumption that $E(e_{X|Y}|Y) = 0$.

Although the OLS estimates are not perfect fits for the regression of $X$ on $Y$, over most of the range of $X$ the linear fit is a decent approximation, and the quadratic fit is very good. We will find that that OLS estimates, though approximate, can be often be serviceable when we impute $X$ conditionally on $Y$.

Having looked at complete bivariate data, we now suppose that some values are *missing at random* (MAR). In the bivariate setting, the MAR assumption means that the probability a value is missing depends only on values that are observed. That is, the probability that $Y$ is missing cannot depend on $Y$, but can depend on $X$ in cases where $X$ is observed. Likewise, the probability that $X$ is missing can depend on observed $Y$ values in cases where $Y$ is observed (Rubin 1976; Heitjan and Basu 1996).

We can impute missing values using several different methods.

## 3.1 Linear regression imputation

The simplest parametric imputation model assumes that the missing values fit a linear regression with normal residuals. We call this *linear regression imputation*.

To understand the technique it is helpful to start with a situation where the assumptions of the imputation model are met. Imagine that $X$ is complete and $Y$ is missing at random (MAR). Then the missing values fit the linear regression in equation (4), and the MAR assumption means that the probability $Y$ is missing is independent of $e_{Y|X}$. In other words, the distribution of $e_{Y|X}$ is the same in the cases with $Y$ observed as in the cases with $Y$ missing, and therefore we can obtain consistent estimates of the regression parameters $\hat{\alpha}_{Yobs|X}, \hat{\beta}_{Yobs|X}, \hat{\sigma}^2_{Yobs|X}$ by simply regressing the $Y$ on $X$ in the cases with $Y$ observed (Anderson 1957; Little 1992; von Hippel 2007). Then to impute missing $Y$ values, we can simulate draws from the regression model in equation (4), with the observed-$Y$ estimates substituted for the parameters:

$$Y_{imp} = \hat{\alpha}_{Yobs|X} + \hat{\beta}_{Yobs|X}Y + e_{Y|X,imp}, \text{ where } e_{Y|X,imp} \sim N(0, \hat{\sigma}^2_{Yobs|X}) \tag{5}$$

An important detail is what observed-$Y$ estimates should be used to run the imputation model. The simplest choice is the OLS estimates: $\hat{\alpha}_{Yobs|X,OLS}, \hat{\beta}_{Yobs|X,OLS}, \hat{\sigma}^2_{Yobs|X,OLS}$. And the estimates that are usually used in imputation models are obtained by adding random variation to the OLS estimates to obtain Bayesian posterior draw (PD) estimates:

$$\hat{\sigma}^2_{Y|X,PD} = \hat{\sigma}^2_{Yobs|X,OLS} \frac{n_{Yobs} - 2}{U}, \text{ where } U \sim \chi^2_{n_{Yobs} - 2 + v_{prior}}$$

$$\begin{bmatrix} \hat{\alpha}_{Y|X,PD} \\ \hat{\beta}_{Y|X,PD} \end{bmatrix} \sim N\left( \begin{bmatrix} \hat{\alpha}_{Yobs|X,OLS} \\ \hat{\beta}_{Yobs|X,OLS} \end{bmatrix}, \hat{V} \begin{bmatrix} \hat{\alpha}_{Yobs|X,OLS} \\ \hat{\beta}_{Yobs|X,OLS} \end{bmatrix} \right) \tag{6}$$

where $n_{Yobs}$ is the number of cases with $Y$ observed, $v_{prior}$ is the prior degrees of freedom, and $\hat{V}[\hat{\alpha}_{Yobs|X,OLS}, \hat{\beta}_{Yobs|X,OLS}]$ is the usual covariance matrix of the OLS estimates, with the estimate $\hat{\sigma}^2_{Y|X,PD}$ substituted for the unknown parameter $\sigma^2_{Y|X}$ (Little and Rubin 1989; Kim 2004).

Like the OLS estimates, the PD estimates are consistent[4], and if $Y$ is regressed on $X$ in the imputed data, the resulting regression estimates are consistent as well, although $\hat{\sigma}^2_{Y|X,PD}$ can be biased in small samples (Kim 2004).

In sum, if $Y$ is the incomplete variable, linear regression imputation yields consistent estimates.

---

[4] To verify this the consistency of the PD estimates, note that, as $n_{Yobs} \rightarrow \infty$, $\hat{\alpha}_{Yobs|X,OLS}, \hat{\beta}_{Yobs|X,OLS}, \hat{\sigma}^2_{Yobs|X,OLS}$ converge to $\alpha_{Y|X}, \beta_{Y|X}, \sigma^2_{Y|X}$, while $(n_{Yobs} - 2)/U$ and $\hat{V}[\hat{\alpha}_{Yobs|X,OLS}, \hat{\beta}_{Yobs|X,OLS}]$ converge to 1 and 0.

If $X$ is the incomplete variable, however, linear regression imputation can be inconsistent. With $X$ incomplete, linear regression imputation assumes that $X$ fits the following equation:

$$X = \alpha_{X|Y} + \beta_{X|Y} Y + e_{X|Y}, \text{ where } e_{X|Y} \sim N(0, \sigma_{X|Y}^2) \tag{7}$$

But in discussing Figure 4 we saw that this equation is incorrect; in particular, $e_{X|Y}$ is non-normal, the conditional expectation of $E(e_{X|Y}|Y)$ is positive for small and large values of $Y$, and the conditional expectation of $X$ is not a linear function of $Y$. Therefore the parameters $\alpha_{X|Y}$ and $\beta_{X|Y}$, to the degree that they are meaningful, cannot be estimated consistently using OLS. And PD estimates that are based on the OLS estimates will be inconsistent as well.

Despite its inconsistency for incomplete $X$, linear regression imputation is a convenient approximation, and we will find out later that its biases are fairly small.

## 3.2   Quadratic regression imputation

*Quadratic regression imputation* is an attempt to improve on linear regression imputation by adding a squared term $Y^2$ to the regression equation for $X$. As we saw in Figure 4, a quadratic term can improve the fit of the regression. We should recognize, however, that the quadratic model is not a perfect specification, and will not fit all data as well as it fits the data in Figure 4. A misspecified quadratic model is vulnerable to extrapolation error. That is, the curve that best fits the observed $X$ values may not be the best fit for the missing $X$ values if the observed and missing $X$ values are in different parts of the distribution.

## 3.3   Imputing within bounds: censoring, truncation, and truncated regression

As in the univariate case, in the bivariate case, we can increase the skew of a normally imputed $X$ by keeping the imputed values within bounds.

The simplest way to bound imputed values is *censoring*. We impute $X$ values under a linear regression imputation model, and then round (censor) out-of-bounds $X$ values up to the boundary value $c$. In our running example, where we impute an exponential variable as though it were conditionally normal, we would round negative imputed values up to zero, which is the lower bound of the exponential distribution.

An alternative approach is *truncation*. Under a simple truncation algorithm, we impute $X$ values under a linear regression imputation model, and then reject any out-of-bounds (e.g., negative) $X$ values and re-impute them until an in-bounds (e.g., positive) $X$ value randomly occurs. This is the approach used by the MI procedure in SAS 9.2, which will terminate if the wait for an in-bounds value is too long. A faster approach is to impute from a truncated distribution using sampling importance resampling (Robert 1995; Raghunathan et al. 2001).

As in the univariate setting, censoring and truncation can improve the shape of the imputed distribution, but censoring and truncation also affect the mean and variance of $X$ in ways that can add bias. And in the bivariate setting, censoring and truncation do not solve the problem that $X$ has been imputed from a misspecified regression model whose parameters are inconsistently estimated. So bounding linear regression imputations has several sources of bias: bias from bounding imputed values, bias from a misspecified regression model, and bias from inconsistent parameter estimates. If we are lucky, these biases will offset each other; if we are unlucky, they will reinforce one another.

A more sophisticated approach is to impute values with a *truncated regression model* (Goldberger 1981),which is implemented by IVEware's BOUNDS statement and by Stata's *mi impute truncreg* command (Raghunathan, Solenberger, and Van Hoewyk 2002; Stata Corp. 2011). Under truncated regression imputation, we do not just truncate the imputed $X$ values. We also estimate the regression parameters under the assumption that the observed $X$ values come from a conditionally normal distribution that has been truncated in the same way. This is akin to the univariate approach that we introduced in Section 2.2, where the effect of truncation was anticipated in estimating the model parameters.

Like the normal regression model, the truncated regression model makes an assumption about the conditional distribution of $X$, and the assumption of a truncated distribution is not correct for many skewed variables. In Figure 4, for example, $X$ follows an exponential distribution, not a truncated conditional normal distribution. The truncated regression model is only an approximation, and the approximation is not necessarily better than the approximation offered by a model without truncation. In fact, Section 2.2 gave us reason to fear that, when the assumptions of a truncated model are violated, the model parameters can be infinite and the estimates can be very sensitive or out-of-bounds.

## 3.4   Transformation

As in the univariate setting, in the bivariate setting it is commonly recommended that skewed incomplete variables be transformed to approximate normality before imputation (Schafer and Olsen 1998, p. 550; Schafer and Graham 2002, p. 167; Allison 2002, p. 39; Raghunathan et al. 2001, pp. 82-83). But transformation of incomplete variables is often unnecessary and can lead to bias.

Again, it is helpful to return to a simple example where $X$ is standard exponential and $Y$ is conditionally normal (see equation (4)). What should we do if $Y$ is missing at random? $Y$ has some skew which it inherits from $X$, so we might be tempted to transform $Y$ to a better approximation of normality. Yet transformation could only hurt the imputations; as we saw in section 3.1, the optimal imputation model is a linear regression that imputes $Y$ on its original scale. Linear regression imputation is appropriate for $Y$ because, although the marginal distribution of $Y$ is skewed, the $X$-$Y$ relationship is linear and the conditional distribution of $Y$ is normal—i.e., the residual $e_{Y|X}$ is normal in the regression of $Y$ on $X$ (see equation (4)). Transformation can only hurt the imputations by introducing nonlinearity and residual nonnormality.

So transformation can only hurt the imputation of $Y$. Transformation can also hurt the imputation of $X$, because transformation introduces curvature into the relationship between $X$ and $Y$. In Figure 4, for example, we have an exponentially distributed $X$, which, as we saw in Section 2.3, can be transformed to a very good approximation of normality by a fourth root transformation $t(X) = \sqrt[4]{X}$ (Hawkins and Wixley 1986). Yet if we use this transformation in a bivariate setting, the imputation model will be misspecified and incompatible with the analysis model. The analysis model is a linear regression that correctly assumes a linear relationship between $Y$ and $X$, yet under transformation the imputation model assumes a linear relationship between $Y$ and $\sqrt[4]{X}$.

As an alternative to transforming $X$ alone, we can transform *all* the variables in the imputation model. For example, if $X$ is exponential, we can transform $X$ to $\sqrt[4]{X}$ and $Y$ to $\sqrt[4]{Y - Min(Y)}$ where $Min(Y)$ is the smallest observed $Y$ value, subtracted to ensure that the value under the radical is zero or more. Here the purpose of transforming $Y$ is not to normalize $Y$, but to put the complete variable $Y$ on a scale that is more nearly linear with respect to the transformed incomplete variable $X$. If both $X$ and $Y$ were incomplete, the choice of transformation would be less straightforward since different transformations would likely be optimal for each variable. And if there were more than two incomplete variables, choosing a suitable transformation could become unwieldy.

# 4   SIMULATION EXPERIMENT

In this section we carry out a simulation experiment to test the bias and efficiency of different methods for imputing a skewed $X$ variable in a bivariate $(X,Y)$ data set. We then extend the simulation to accommodate a third variable $Z$.

## 4.1   Bivariate design

In each simulated bivariate dataset, $Y$ fits a linear regression on $X$:

$$Y = \alpha_{Y|X} + \beta_{Y|X}X + e_{Y|X}, \text{ where and } e_{Y|X} \sim N(0, \sigma_{Y|X}^2) \tag{8}$$

We hold the slope and intercept constant at $\alpha_{Y|X} = \beta_{Y|X} = 1$ since changing these values to any nonzero value would be equivalent to a trivial shifting or rescaling of the axes. We also hold the sample size constant at $n$=100 observations, so that we can avoid being distracted by small sample biases[5] and focus on the asymptotic properties of the imputation methods.

The experiment independently manipulates four factors:

1. The first factor is the *distribution* of $X$. We let $X \sim \chi_\nu^2$ be a chi-squared variable where the degrees of freedom $\nu$ determines the mean $E(X) = \nu$, the variance $V(X) = 2\nu$, and the coefficient of skewness $Skew(X) = \sqrt{8/\nu}$. We set the degrees of freedom to four different levels $\nu = 1,2,4,8$ so that the skew ranges from mild (1) to severe ($\sqrt{8} \approx 2.8$). Note that with $\nu = 2$ the chi-square distribution reduces to the standard exponential distribution that we have used in all our examples up to this point.

---

[5] Small sample biases can afflict multiple imputation estimates even when the data are normal and the imputation model is correctly specified (Kim 2004; Demirtas and Hedeker 2008; von Hippel 2012).

2. The second factor is the strength of the *X-Y* relationship as measured by the regression's *coefficient of determination* $R^2_{Y|X}$. We set the coefficient of determination to four different levels $R^2_{Y|X} = .1, .3, .5, .7, .9$ by first choosing the desired value of $R^2_{Y|X}$ and then setting $\sigma^2_{Y|X} = \beta^2_{Y|X} Var(X)(1 - R^2_{Y|X})/R^2_{Y|X}$.

3. The third factor is the *location* of the missing *X* values. From each complete dataset, we delete half the *X* values in three MAR patterns:

    a. To let values be missing completely at random (MCAR), we delete *X* values with a constant probability of ½.

    b. To favor deletion of large *X* values—i.e., values in the *tail* of the *X* distribution— we delete *X* values with probability $\hat{F}(Y)$ where $\hat{F}(Y)$ is the empirical cumulative distribution of *Y*.

    c. To favor deletion of small *X* values—i.e., values near the *peak* of the *X* distribution—we delete *X* values with probability $1 - \hat{F}(Y)$.

    Note that we do not vary the *fraction* of missing *X* values, since it is clear that changing the fraction of missing values would not change the relative performance of the imputation methods. Deleting a larger or smaller fraction of values would just make the differences between the imputation methods more or less consequential.

4. The fourth factor is the *method of imputation*. We use the seven different methods described in Section 3:

    a. *Linear regression imputation*, where *X* is regressed on *Y*.
    b. Linear regression imputation with imputed values *censored* below $c = 0$.
    c. Linear regression imputation with imputed values *truncated* below $c = 0$.
    d. *Quadratic regression imputation*, where *X* is regressed on *Y* and $Y^2$.
    e. The *transform X* method: linear regression imputation where $\sqrt[4]{X}$ is regressed on *Y*. before imputation. For a chi-squared variable *X*, $\sqrt[4]{X}$ is very close to normal for any degrees of freedom $\nu$ (Hawkins and Wixley 1986).
    f. The *transform all* method: linear regression imputation where $\sqrt[4]{X}$ is regressed on $\sqrt[4]{Y - Min(Y)}$. Here $Min(Y)$ is the smallest *Y* value in each simulated dataset, subtracted to ensure that the value under the radical is nonnegative.
    g. *Truncated regression imputation*, with the lower truncation point at $c = 0$.

For each imputation method, we imputed each incomplete dataset five times. In each of the five imputed datasets, we calculated the mean, standard deviation, and skew of *X*, and we fit an OLS regression of *Y* on *X*. We then averaged estimates from the five imputed datasets to yield a single set of multiple imputation (MI) estimates for each incomplete dataset and imputation method.

All the imputation methods used the MI procedure in SAS 9.2, except for the truncated regression method, which used the *mi impute truncreg* command in Stata 12. Both SAS and Stata displayed some technical limitations. These limitations are not central to the results, but are worth describing briefly, as follows:

- In truncating normal imputations, SAS's MI procedure uses a simple rejection algorithm[6], rejecting negative values and re-imputing them iteratively until a positive value randomly occurs. The wait can be long, and in about 8% of datasets SAS terminated the attempt after 100 iterations. We re-imputed these datasets with a lower truncation point of $c = -6$, which means that the truncation was not as severe as we wanted it to be. We could have maintained a truncation point of $c = 0$ if SAS truncated imputed values by sampling importance resampling, which is used by IVEware (Raghunathan et al. 2001, 2002). Sampling importance resampling is faster than simple rejection and ensures that imputed values are within bounds (Robert 1995).

- In transforming imputed variable, SAS's MI procedure[7] refuses to inverse-transform any values that could not have been obtained by transformation. In imputing $X$, for example, we transformed the observed $X$ values to $X_{obs,t} = \sqrt[4]{X_{obs}}$ and then supplemented the observed values with normal imputations $X_{imp,t}$. Naturally, the imputations included some negative values. When we tried to invert the transformation by calculating $X_{imp} = X_{imp,t}^4$, SAS terminated on encountering negative values of $X_{imp,t}$ because a fourth root cannot be negative. We got around this problem by transforming and inverse-transforming the variable outside of the MI procedure.

- Stata's implementation of truncated regression imputation is quite slow, taking about 24 hours to impute all the experimental datasets. (SAS imputed them in minutes.) It is not clear whether this is an inherent problem with truncated regression, or a problem in Stata's implementation of it. In addition to being slow, the truncated regression model failed to converge, in about 1% of incomplete datasets—typically datasets with low $\nu$ and low $R_{YX}^2$. While again this could be an implementation issue, we suspect that failure to converge resulted from the truncated regression model's inherently poor fit to a chi-square variable $X$ with low $\nu$. Recall that, even in the univariate setting (Section 2.2), we found that fitting a truncated normal model to a chi-square variable with $\nu = 2$ could lead to infinite parameters and unstable parameter estimates. After failures to converge we re-fit the truncation regression, successfully, by moving the lower truncation point to $c = -1$.

## 4.2  Illustrative results

Figure 5 uses all the imputation methods to impute a simulated dataset with $\nu = 2$, $R_{YX}^2 = .3$, and $X$ values MCAR. The true regression line—that is, the line that fits the complete-data population—is shown for reference.

<center>←Figure 5 <b>near here</b>→</center>

Clearly some methods handle these data better than others. Under linear regression imputation, the imputed values, although more symmetrically distributed than the observed values, nevertheless fit well around the true regression line. Censoring or truncating the linear regression imputations does not change the fit very much; censoring and truncating simply move the leftmost values a little to the right, and this has little influence since the points with the most influence on the regression line are on the far right, in the tail of the $X$ distribution. Quadratic regression imputation produces fairly similar results, although the imputed $X$ values are a little more dispersed than they are under linear regression imputation.

---

[6] The rejection algorithm is invoked by the MINIMUM and MAXIMUM options in the MI procedure.
[7] The MI procedure implements transformation and inverse transformation via the TRANSFORM option.

The transform $X$ and transform all methods have serious trouble with these data, because they curve the relationship between $X$ and $Y$. In the tail, which has the most influence on the regression, all the imputed points lie below the true regression line. These points will negatively bias the estimated slope.

The truncated regression has even worse problems, with nearly all the imputed points in influential positions and below the true regression line. Evidently the truncated regression model can be a poor fit to observed data that are not actually truncated, and very unrealistic parameter estimates can result. Recall that the truncated model can produce infinite parameter values even in a simple univariate setting (Section 2.2).

Figure 6 gives even more serious examples of bad imputations obtained from the truncated regression model. In one example, the imputed $X$ values have a negative relationship to $Y$, even though the relationship in the observed data is clearly positive. Some imputed $X$ values have values of 200 or greater, even though all the observed $X$ values are less than 11. In one example, nearly all the imputed values are negative, even though the point of this truncated regression model is to produce positive imputations. The presence of negative imputations, though extremely rare, suggests implementation problems that go beyond the basic difficulty of fitting a truncated model to non-truncated data.

←Figure 6 **near here**→

## 4.3   Comprehensive results

The full bivariate simulation experiment yielded 36,000 sets of MI estimates—100 estimates for each of 360 different experimental conditions (4 $\nu$ values ✕ 5 $R_{YX}^2$ values ✕ 3 missing value patterns ✕ 6 imputation methods). Within each experimental condition, we calculated the mean and standard deviation of the 100 MI estimates in order to estimate the expectation and standard error of each imputed-based estimator under each experimental condition. From the expectation and standard error, we calculated the bias and root mean squared error (RMSE). Finally we divided the bias and RMSE by the true value of the estimand in order to get the *relative bias* and *relative RMSE*—that is, the bias and RMSE expressed as a percentage of the true value of the estimand.

Table 1 summarizes the relative bias and relative RMSE of each imputation method, averaged across all the experimental conditions. Results are shown for seven different estimands. The first three estimands describe the marginal distribution of $X$ in terms of the mean, standard deviation, and coefficient of skewness. The remaining four estimands describe the regression of $Y$ on $X$ in terms of the intercept, slope, residual standard deviation, and coefficient of determination.

For all but one of the estimands, linear regression imputation yields estimates whose average relative bias is fairly small, often close to zero. Censoring or truncating the linear regression imputations yields similar results, with a little more bias for some estimands and a little less for others.[8] All the other methods have more serious biases. The transformation methods have the worst biases for the regression slope and intercept, and the truncated regression method has the worst biases for the mean and standard deviation of $X$. The biases of the truncated regression method can be enormous, exceeding four hundred thousand percent for some parameters, because the method occasionally imputes wild outliers like those in Figure 6.

In short, for the estimands that are emphasized in most social research—means, standard deviations, and regression parameters—linear regression imputation, with or without censoring or truncation, can often produce reasonable results. The other imputation methods are no better for these estimands, and sometimes much worse.

This is not to say that linear regression imputation is good for every estimand—it isn't. In fact, linear regression imputation does a very poor job of estimating parameters that reflect distributional shape. In estimating the coefficient of skewness, for example, linear regression imputation had an average relative bias of –52%—that is, the estimated skew was less than half the true skew, on average. With one exception, the other methods don't estimate skew very well either; they are less biased than linear regression imputation, but still have positive or negative biases exceeding 20%. The one method that estimates skew with little bias is the transform-all method, but we cannot recommend that method because it has serious biases in estimating the regression parameters.

In the Appendix, Table A-1 summarizes the simulation results in more detail, breaking them down across different levels of the manipulated factors. The most striking result is that the differences of the methods are largest when missing values are in the tail. This makes sense because the tail values have the most influence in estimating the regression line. Another striking result is that the biases of the transformation methods are worst when $X$ is highly skewed— which is exactly when one would be most tempted to use transformation.

The biases of the different methods vary from one simulated condition to another, and there are circumstances where transformation or quadratic regression or truncated regression gives the best results. It is tempting to imagine that you could obtain good results by picking and choosing different methods to suit different circumstances, but that it is a dangerous game with only small benefits for guessing right, and large penalties for guessing wrong.

The safest approach is to use a method that tends to have small biases in a wide variety of settings, and by that criterion the best choice for bivariate data is linear regression imputation, with or without truncation or censoring.

---

[8] As Figure 2 would lead us to expect, censoring and truncation increase the mean and reduce the standard deviation. The effect of truncation is less than it should be because the software would not allow us to truncate all the datasets at $c$=0.

## 4.4   Trivariate extension

We now extend the simulation to a regression where there are three variables: a complete dependent variable $Z$ and two independent variables, one complete ($Y$), and one incomplete ($X$). To do this, we simply keep the $X$ and $Y$ variables from the simulation in Section 4, and add a $Z$ variable that fits a linear regression on $X$ and $Y$, with normal residuals:

$$Z = \alpha_{Z|XY} + \beta_{Z|X}X + \beta_{Z|Y}Y + e_{Z|XY}, \text{ where } e_{Z|XY} \sim N(0, \sigma^2_{Z|XY}) \tag{9}$$

The manipulated factors are the same as before, with only necessary adjustments:

1. The first factor is the distribution of $X$. Again $X \sim \chi^2_\nu$ was a chi-squared variable with degrees of freedom $\nu = 1,2,4,8$.

2. The second factor is the squared correlation $\rho^2_{XY}$ between $X$ and $Y$. This is the factor that we manipulated in the bivariate simulation, except that in the bivariate simulation $\rho^2_{XY}$ was interpreted as the coefficient of determination $R^2_{Y|X}$ for the regression of $Y$ on $X$.

3. The third factor is the location of missing $X$ values. These patterns are defined in exactly the same way as in the bivariate setting. In the MCAR scenario, $X$ values are deleted with probability ½; in the tail scenario, $X$ values are deleted with probability $\hat{F}(Y)$; and in the peak scenario, $X$ values are deleted with probability $1 - \hat{F}(Y)$. Note that $Z$ has no effect on whether $X$ is missing, and that both $Y$ and $Z$ were complete.

4. The fourth factor was the imputation method. The methods were the same as in the bivariate simulation, except that $X$ was imputed conditionally on $Z$ as well as Y. In the transform all method, $X$ was transformed to $\sqrt[4]{X}$, $Y$ was transformed to $\sqrt[4]{Y - Min(Y)}$, and $Z$ was transformed to $\sqrt[4]{Z - Min(Z)}$, where $Min(Y)$ and $Min(Z)$ are the minimum $Y$ and $Z$ values in the imputed dataset, subtracted to ensure that the value under the radical is nonnegative.

To avoid further complicating the experiment we hold constant certain parameters of the regression of $Z$ on X and Y. Those parameters are the slope and intercepts, which are held constant at $\alpha_{Z|XY} = \beta_{Z|X} = \beta_{Z|Y} = 1$, and the coefficient of determination, which is held constant at $R^2_{Z|XY} = .5$ by setting $\sigma^2_{Z|XY} = V(\beta_{Z|X}X + \beta_{Z|Y}Y)$.

Table 2 summarizes the relative bias and relative RMSE of each imputation method for estimands relating to the marginal distribution of $X$ and the regression of $Z$ on $X$ and $Y$. Table 2 presents average results; in the Appendix, Table A-2 breaks the results down by each manipulated factor.

The basic results of the 3-variable simulation are similar to those of the 2-variable simulation.

In estimating the regression parameters, and in estimating the mean and standard deviation of $X$, linear regression imputation gives perhaps the best results overall, with relative biases of 5 percent or less. Quadratic regression imputation gives similar results, as does censoring or truncating the linear regression imputations. The transformation methods give more biased regression parameters, and the truncated regression method occasionally imputes very large outliers and so gives seriously biased estimates of the mean and standard deviation.

In estimating the skew, all the methods give highly biased results except for the transform all method, which cannot be recommended since it gives the most biased estimates for the regression. Again, the differences among the imputation methods are much more consequential if values are imputed in the tail rather than the peak of the $X$ distribution (Table A-2b).

In the 3-variable simulation, the slope of $Z$ on $Y$ has bias, which is opposite in direction to the bias of the slope of $Z$ on $X$. Bias in the slope of $Z$ on $Y$ occurs despite the fact that neither $Z$ nor $Y$ has any imputed values. The reason for this bias is that $Y$ is correlated with $X$, so any bias in the slope of $X$ engenders a compensating bias in the slope of $Y$. Table A-2b in the Appendix confirms that the bias in the slope of $Y$ is greatest when the squared correlation $\rho_{XY}^2$ between $X$ and $Y$ is large. Conceptually, the bias is similar to the bias that would occur if $X$ were omitted or measured poorly.

An initially surprising result of the 3-variable simulation is that the regression slopes estimated by the transformation methods, despite having the most bias, sometimes have smaller RMSE than the regression slopes estimated by other methods. The presumed reason for this is that the transformation methods, because they use misspecified imputation models, reduce the correlation between $Y$ and the imputed $X$, and this reduced correlation yields smaller standard errors when $X$ and $Y$ are used to predict $Z$. Under some circumstances, the reduction in standard error can more than make up for the increase in bias.

Is the potential for reduction in RMSE a reason to use the transformation methods? No. First, reduced RMSE in the regression slope is limited to the simulations where the correlation between $X$ and $Y$ is very, very high. Table A-2b in the Appendix shows reduced RMSE in the simulation where $\rho_{XY}^2 = .9$ (i.e., $\rho_{XY} \approx .95$), but no reduction in RMSE if $\rho_{XY}^2 = .7$ (i.e., $\rho_{XY} \approx .84$). Second, because RMSE was reduced by accident, we have no theory for exactly when a similar reduction in RMSE might be expected outside the simulation. Analysts who use the transformation methods are risking severe bias, with only a speculative hope of reduced RMSE. Analysts who are willing to trade some bias for a reduction in RMSE would be better advised to use a more principled method with stronger theoretical foundations, such as ridge regression (Hoerl and Kennard 1970; Muniz and Kibria 2009).

# 5   APPLIED EXAMPLES

Authors often use incomplete skewed variables in applied research. In this section we discuss two examples, and evaluate the decisions that were made by the authors.

## 5.1    Analysis of female legislative candidates

In a cross-national study, Kunovich and Paxton (2005) pointed out a strong bivariate relationship between two percentages that vary across the world's $n$=171 countries: the percentage of parliamentary *legislators* who are female ($Y$) and the percentage of parliamentary *candidates* who were female ($X$). The regression of $Y$ on $X$ is important because a slope of less than one suggests that female candidates lose more often than they win. $Y$ was complete and skewed to the right, with a coefficient of skewness of 1.4; more concretely, in most countries $Y$ was less than 11% of the candidates were female, but in five north European countries $Y$ exceeded 30%.  The skew of $X$ is harder to describe from the observed data, since $X$ was missing for more than half (99) of the countries.

Following my advice, the authors imputed missing $X$ values using the transform $X$ method. It now appears, in light of results in this paper, that my advice was misguided and could have caused substantial bias if missing $X$ values had been primarily in the tail. Fortunately, $X$ was missing primarily in the peak, and in this situation—with a skew of 1.4, values missing from the peak, and $R^2_{YX} = .63$—simulation suggests that the transformation method is nearly unbiased, while imputation without transformation has a bias of about −12% (Appendix A). Note that the authors used a milder transformation than the fourth-root transformation evaluated in our simulation. In particular, the authors used the arcsine transformation $\sin^{-1}\sqrt{X}$, which is practically indistinguishable from $\sqrt{X}$ for the range of $X$ values in their data.

Reanalyzing the authors' data without control variables, I found that the estimated slope of $Y$ on $X$ was .60 with the authors' transformation, and .55 without transformation. My simulations suggest that the authors' estimate is closer to the truth, but the difference is not large either way.

## 5.2    Analysis of body mass index (BMI)

In a longitudinal study of $n$=358 children, von Hippel, Nahhas, and Czerwinski (2012) estimated percentiles for change in body mass index (BMI), from age 3½ to age 18 years. BMI was measured every six months, but some measurements were missing. Our longitudinal imputation model was more complicated than the models considered in this paper, but the model still assumed that the incomplete variables were conditionally normal. Skew in BMI remained an important challenge, especially since the skew of the BMI distribution increases as children grow older.

As one of the authors, I knew that linear regression imputation could estimate the conditional mean of a skewed variable with little bias, but this was not reassuring since in this study we sought to estimate the percentiles instead of the mean. With or without transformation, it proved difficult to find an imputation model that gave plausible estimates for extreme tail percentiles such as the 90[th] or 95[th].

In the end, we sidestepped the issue by imputing BMI increments—that is, changes in BMI from one measurement occasion to the next. Although BMI is skewed, the distribution of BMI increments is approximately symmetric and has only slightly more kurtosis than a normal distribution.

# 6 CONCLUSION

On the whole, the simulation results suggest that when an incomplete variable has skew, linear regression often gives reasonable, though not unbiased, estimates for the quantities most commonly estimated in social science—namely, means, standard deviations, and linear regressions (see also Demirtas et al. 2008). *Ad hoc* modifications of linear regression—through censoring, truncation, or transformation—rarely do much to improve the estimates, and in fact can make the estimates much worse.

Although the normal regression method is fairly good for estimating means, variances, and regressions, it can do a poor job of estimating quantities that depend on distributional shape. Such quantities include the coefficient of skewness, the percentiles, and quantities based on percentiles such as the Gini coefficient.

Although our discussion has been confined to normal imputation, we note that similar biases would be expected if estimates were obtained, without imputation, by applying maximum likelihood to the incomplete data under an assumption of normality. Maximum likelihood can be asymptotically biased when its distributional assumptions are violated. For example, in Section 3 we discussed a situation where OLS estimates for the regression of $X$ on $Y$—which are maximum likelihood estimates if $X$ is conditionally normal—are biased for a skewed $X$.

To improve estimation from incomplete skewed variables, it would be helpful to have imputation methods that do not assume normality. One option is the distribution-free approach of imputing missing values by resampling values from similar cases. Variants of this idea are called hot-deck imputation, the approximate Bayesian bootstrap, or predictive mean matching (for a review, see Andridge and Little 2010). For example, in bivariate data $(X,Y)$ with $Y$ complete and $X$ missing at random, one would fill in missing $X$ values with observed $X$ values resampled from cases with similar values of $Y$.

While imputation by resampling is initially attractive, it can work poorly when observed values are sparse in one part of the distribution. For example, suppose that $X$ is missing if and only if $Y<0$. If we wish to impute the missing $X$ values in cases with $Y<0$, we have to resample $X$ values from cases with $Y\geq0$—and this inevitably leads to bias. (See Allison 2000 for a similar but slightly more complicated example.) To take a less-artificial example from applied research: in our study of BMI growth among children, the very heaviest children had a number of missing measurements. These measurements could not be imputed by resampling, unless we were willing to resample BMIs from much lighter children.

A perhaps more promising idea is to model and impute non-normal variables using flexible non-normal distributions that can take a variety of shapes—such as Tukey's *gh* distribution (He and Raghunathan 2006), the Weibull or beta density (Demirtas and Hedeker 2008), the generalized lambda distribution, or Fleishman power polynomials (Demirtas 2010). These approaches are currently in the development stage. Initial evaluations suggest that they can mimic very well the shape of many non-normal distributions, and that they preserve the relationships among variables at least as well as normal imputation methods (Bondarenko and Raghunathan 2007; He and Raghunathan 2006, 2009, 2012; Demirtas 2010; Demirtas and Hedeker 2008). We look forward to these flexible methods being available in software so that they can be used in applied research and evaluated further.

# TABLES

*Table 1*. Results of 2-variable simulation. Main effect of imputation method, averaged across other factors.

| Imputation method | Marginal distribution of incomplete X | | | | | | Regression of complete Y on incomplete X | | | | | | | |
|---|---|---|---|---|---|---|---|---|---|---|---|---|---|---|
| | Mean | | SD | | Skew | | Slope | | Intercept | | Residual SD | | $R^2_{Y|X}$ | |
| | Rel. bias | Rel. RMSE | Rel. bias | Rel. RMSE | Rel. bias | Rel. RMSE | Rel. bias | Rel. RMSE | Rel. bias | Rel. RMSE | Rel. bias | Rel. RMSE | Rel. bias | Rel. RMSE |
| Linear regression | -6% | (15%) | -3% | (22%) | -52% | (57%) | 1% | (26%) | 15% | (95%) | 0% | (12%) | 3% | (37%) |
| Linear regression, censored | -1% | (13%) | -8% | (20%) | -37% | (46%) | 5% | (27%) | -1% | (93%) | 1% | (12%) | 0% | (35%) |
| Linear regression, truncated | 2% | (16%) | -9% | (21%) | -38% | (46%) | 4% | (28%) | -6% | (96%) | 2% | (12%) | -5% | (35%) |
| Quadratic regression | 0% | (16%) | 6% | (25%) | -31% | (57%) | -11% | (31%) | 37% | (113%) | 1% | (13%) | 3% | (41%) |
| Truncated regression | 404936% | (4064963%) | 417393% | (4181515%) | -29% | (43%) | -8% | (28%) | 18% | (102%) | 0% | (11%) | 5% | (39%) |
| Transform X | 5% | (21%) | 27% | (62%) | 21% | (48%) | -17% | (30%) | 47% | (106%) | 14% | (21%) | -11% | (37%) |
| Transform all | -5% | (16%) | -1% | (30%) | -5% | (29%) | -20% | (31%) | 74% | (118%) | 25% | (32%) | -31% | (50%) |

*Table 2.* Results of 3-variable simulation. Main effect of imputation method, averaged across other factors.

| | Marginal distribution of incomplete X | | | | | | Regression of complete Z on incomplete X and complete Y | | | | | | | | | |
|---|---|---|---|---|---|---|---|---|---|---|---|---|---|---|---|---|
| | Mean | | SD | | Skew | | Slope of X | | Slope of Y | | Intercept | | Residual SD | | $R^2_{Y|X}$ | |
| Imputation method | Rel. bias | Rel. RMSE | Rel. bias | Rel. RMSE | Rel. bias | Rel. RMSE | Rel. bias | Rel. RMSE | Rel. bias | Rel. RMSE | Rel. bias | Rel. RMSE | Rel. bias | Rel. RMSE | Rel. bias | Rel. RMSE |
| Linear regression | -5% | (14%) | -2% | (22%) | -51% | (57%) | -2% | (65%) | 4% | (42%) | -1% | (157%) | -1% | (8%) | 2% | (17%) |
| Linear regression, censored | -1% | (13%) | -7% | (19%) | -37% | (45%) | 2% | (66%) | 4% | (40%) | -18% | (155%) | -1% | (8%) | 2% | (17%) |
| Linear regression, truncated | 2% | (15%) | -8% | (20%) | -37% | (46%) | -1% | (65%) | 6% | (41%) | -25% | (161%) | 0% | (8%) | 1% | (17%) |
| Quadratic regression | 0% | (15%) | 7% | (25%) | -29% | (57%) | -13% | (64%) | 7% | (41%) | 9% | (155%) | -1% | (8%) | 2% | (17%) |
| Truncated regression | 78933932% | (607613905%) | 71427514% | (546826216%) | -27% | (42%) | -6% | (64%) | 2% | (40%) | -4% | (157%) | -1% | (8%) | 2% | (17%) |
| Transform X | 6% | (20%) | 30% | (63%) | 24% | (51%) | -27% | (57%) | 18% | (37%) | 3% | (142%) | 0% | (7%) | 0% | (17%) |
| Transform all | -4% | (16%) | 0% | (32%) | -5% | (30%) | -31% | (60%) | 22% | (41%) | 11% | (146%) | 1% | (8%) | -1% | (17%) |

**FIGURES**

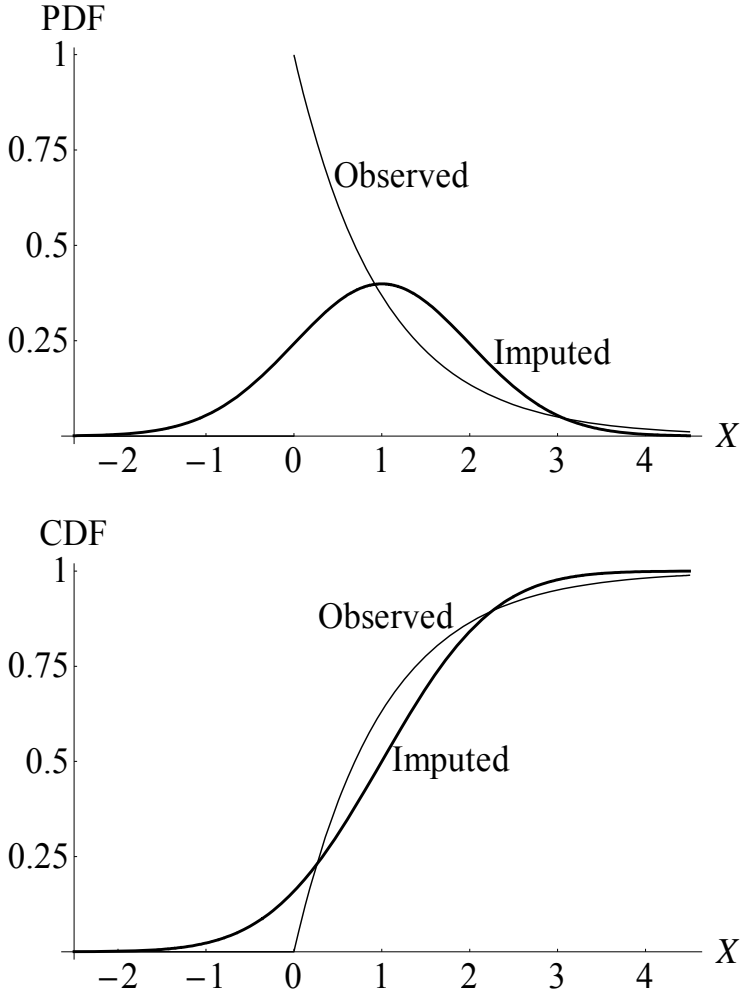

*Figure 1.* Distribution of a standard exponential observed variable and a normal imputed variable that has the same mean and variance.

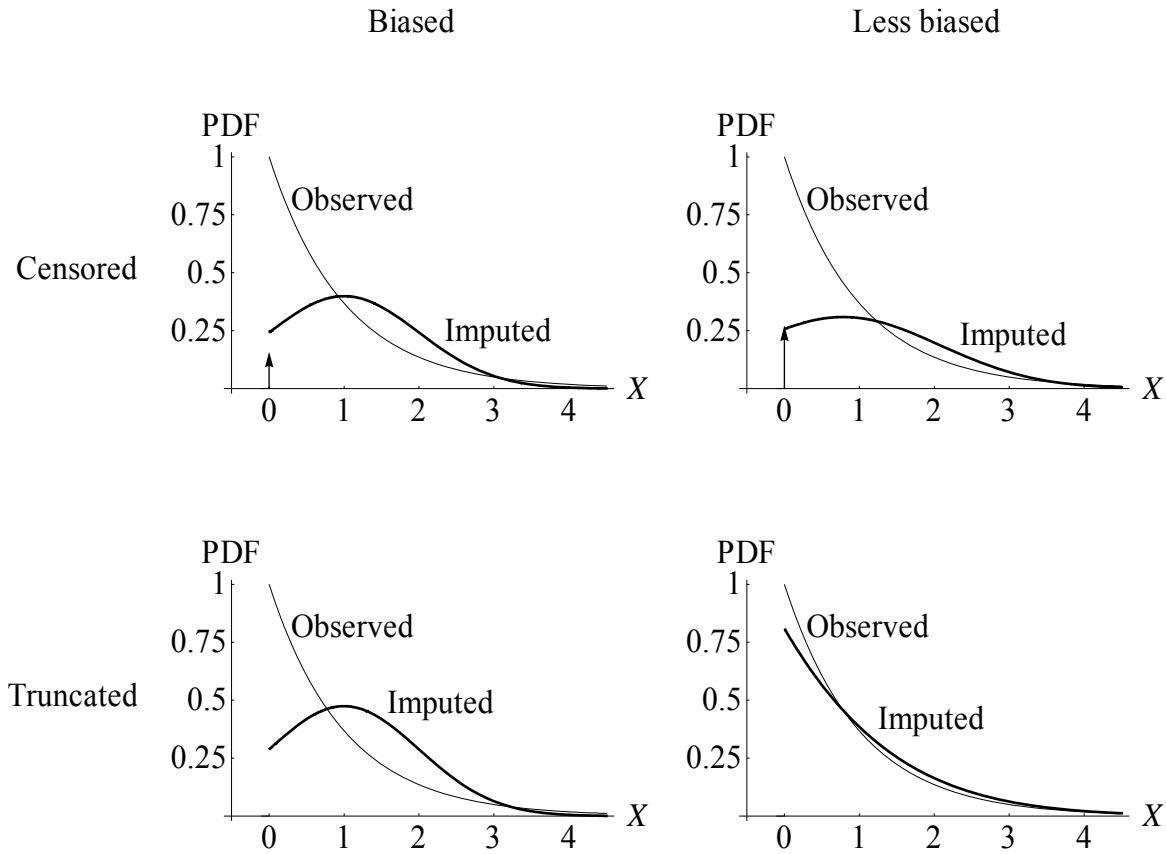

*Figure 2.* Biased and less-biased ways of imputing a censored or truncated normal variable to match the distribution of an observed exponential variable.

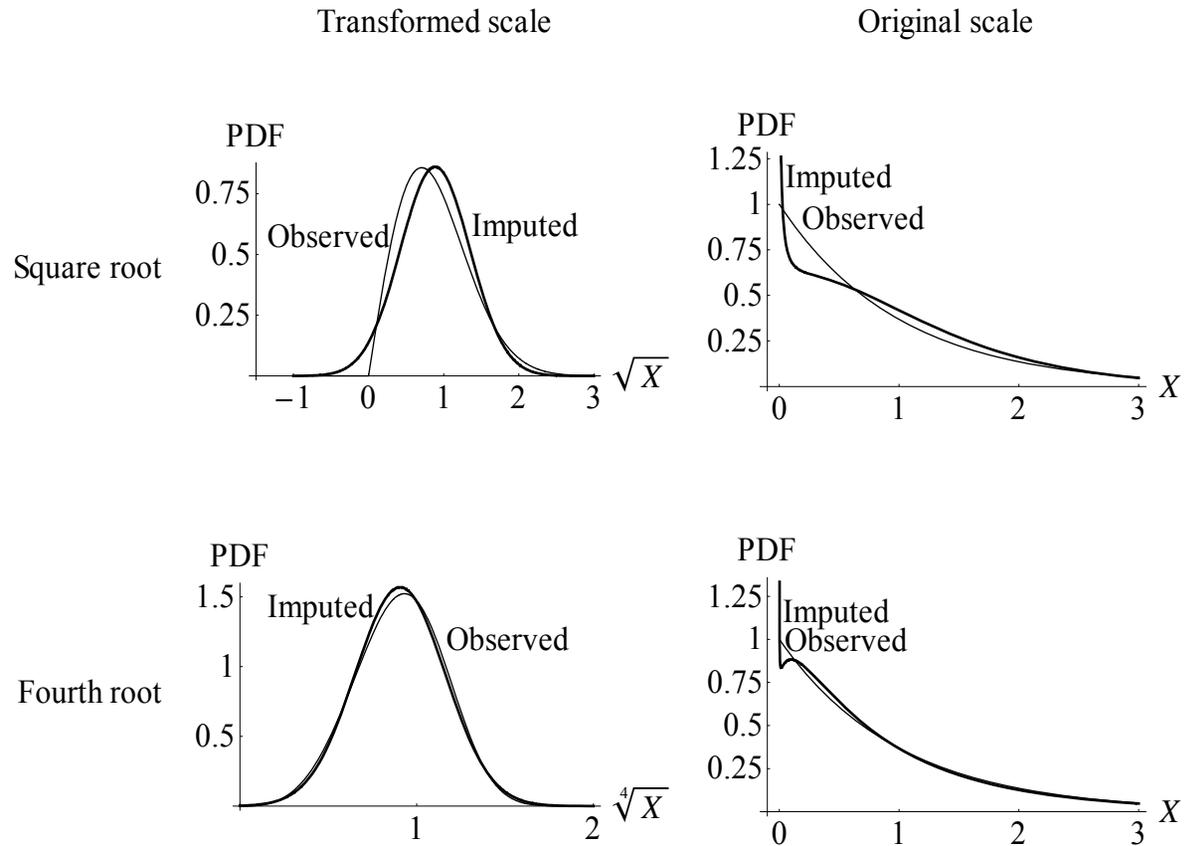

*Figure 3.* Normal imputation of an observed exponential variable under square-root and fourth-root transformation.

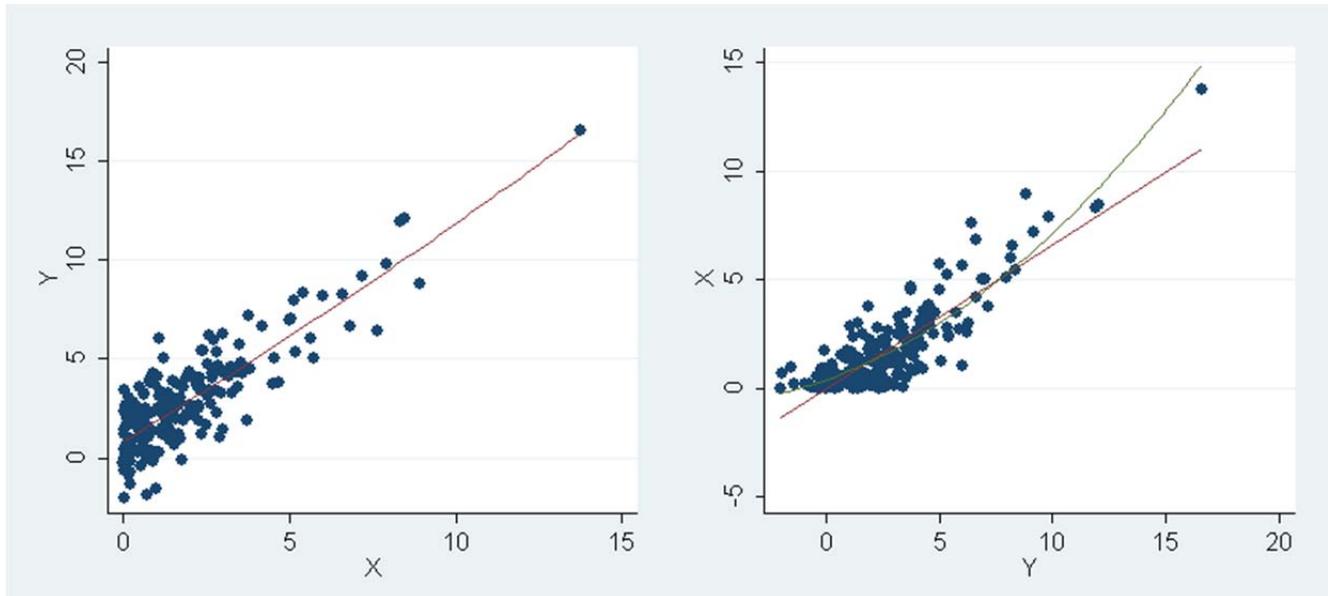

*Figure 4. X* is standard exponential and *Y* is conditionally normal. The left panel shows a scatterplot for the regression of *Y* on *X*, with an OLS line. The right panel shows a scatterplot for the regression of *X* on *Y*, with an OLS line and an OLS quadratic.

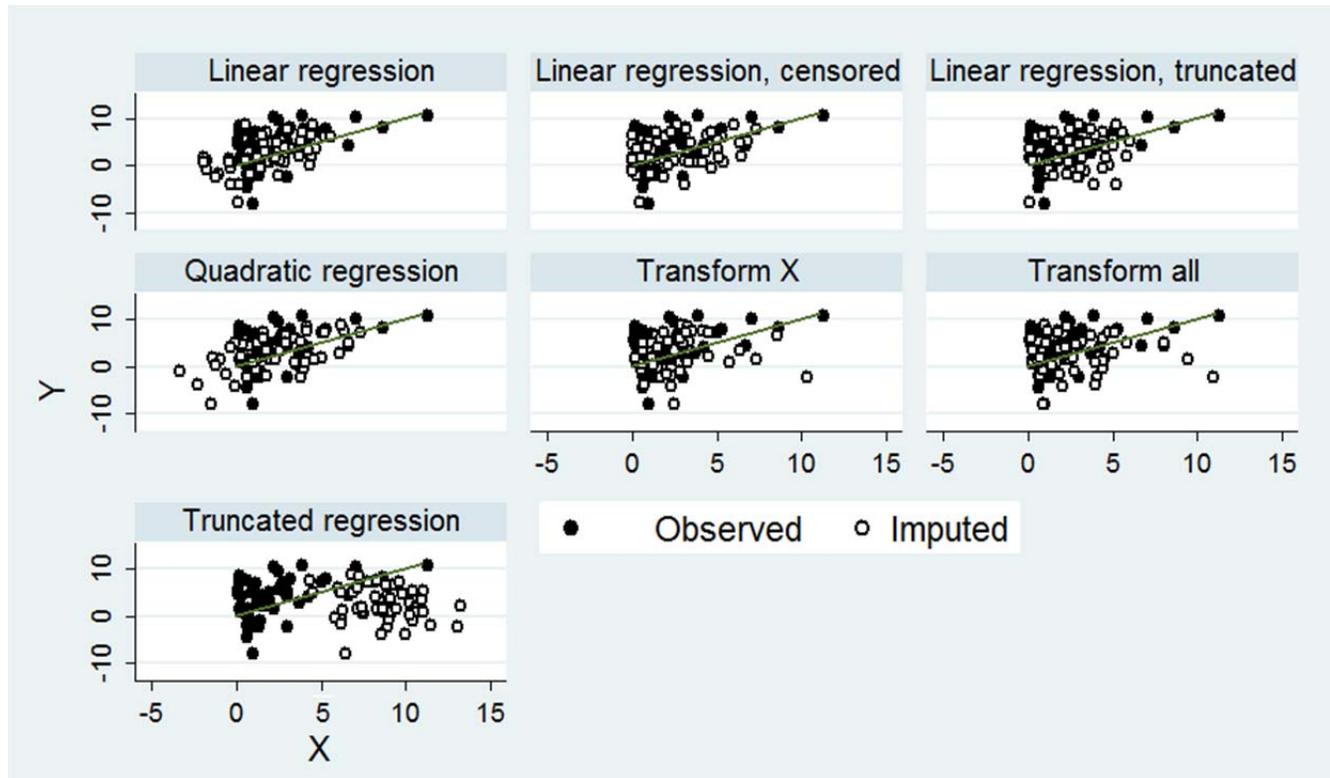

*Figure 5.* Different imputation methods when the incomplete variable *X* is standard exponential with values missing completely at random, and the complete variable *Y* is conditionally normal.

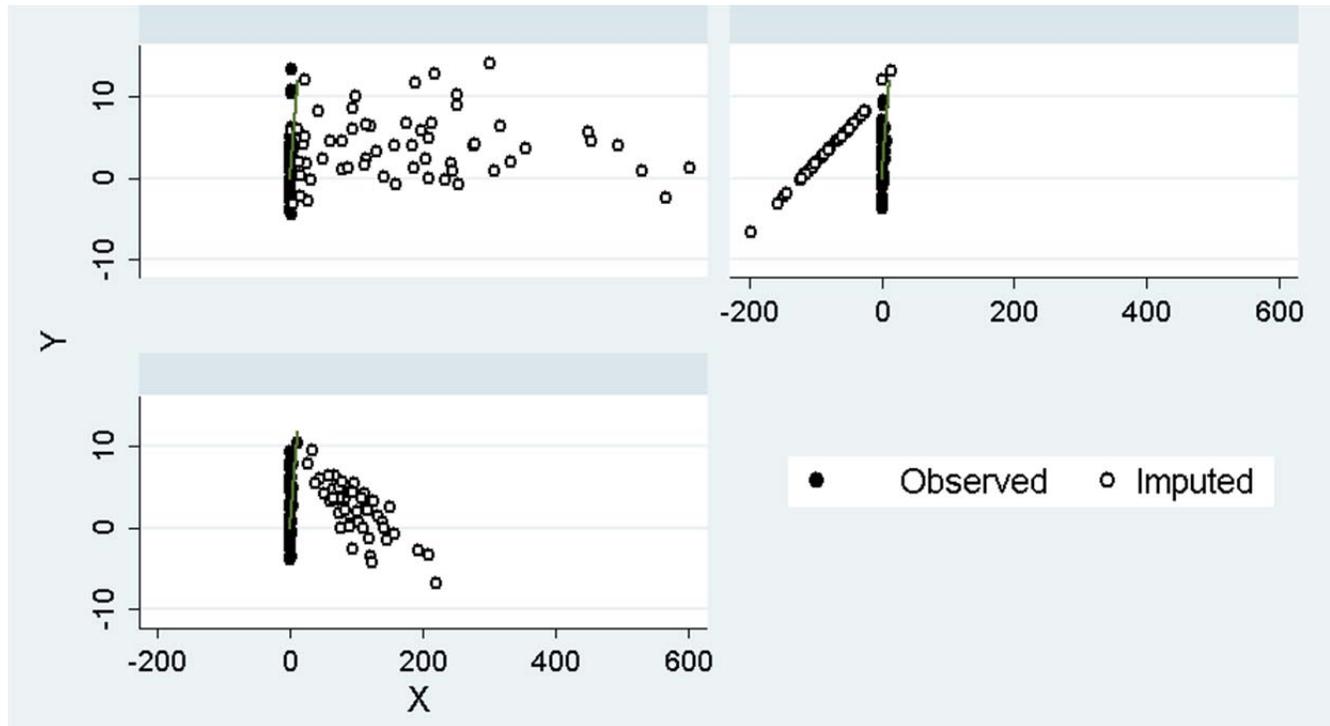

*Figure 6.* Anomalous imputed datasets generated by truncated regression imputation in Stata 12.

# A. APPENDIX

*Table A-1*. Results of 2-variable simulation, by manipulated factors

*Table A-1*a. By degrees of freedom in the *X* distribution

| | | Marginal distribution of incomplete X | | | | | | Regression of complete Y on incomplete X | | | | | | | |
|---|---|---|---|---|---|---|---|---|---|---|---|---|---|---|---|
| | | Mean | | SD | | Skew | | Slope | | Intercept | | Residual SD | | $R^2_{Y|X}$ | |
| ν | Imputation method | Rel. bias | Rel. RMSE | Rel. bias | Rel. RMSE | Rel. bias | Rel. RMSE | Rel. bias | Rel. RMSE | Rel. bias | Rel. RMSE | Rel. bias | Rel. RMSE | Rel. bias | Rel. RMSE |
| 1 | Linear regression | -13% | (25%) | -6% | (31%) | -55% | (58%) | 7% | (35%) | 9% | (38%) | 1% | (14%) | -4% | (55%) |
| | Linear regression, censored | -1% | (21%) | -16% | (28%) | -35% | (42%) | 14% | (35%) | -9% | (35%) | 3% | (13%) | -7% | (56%) |
| | Linear regression, truncated | 3% | (28%) | -14% | (31%) | -43% | (48%) | 11% | (38%) | -10% | (44%) | 4% | (15%) | -9% | (59%) |
| | Quadratic regression | 0% | (26%) | 8% | (37%) | -32% | (55%) | -13% | (39%) | 14% | (43%) | 2% | (15%) | -5% | (56%) |
| | Truncated regression | 1612395% | (1620182396%) | 1658954% | (166505097%) | -32% | (49%) | -21% | (39%) | 31% | (57%) | 2% | (13%) | -6% | (57%) |
| | Transform X | 13% | (42%) | 62% | (144%) | 15% | (42%) | -24% | (37%) | 26% | (42%) | 22% | (28%) | -24% | (50%) |
| | Transform all | -6% | (27%) | 7% | (58%) | 3% | (31%) | -31% | (41%) | 38% | (49%) | 35% | (41%) | -48% | (65%) |
| 2 | Linear regression | -6% | (16%) | -2% | (24%) | -51% | (56%) | 2% | (26%) | 10% | (60%) | 0% | (13%) | -1% | (55%) |
| | Linear regression, censored | -1% | (14%) | -9% | (21%) | -33% | (42%) | 7% | (26%) | -9% | (57%) | 1% | (12%) | -3% | (55%) |
| | Linear regression, truncated | 4% | (17%) | -9% | (22%) | -36% | (45%) | 5% | (28%) | -14% | (65%) | 2% | (12%) | -6% | (56%) |
| | Quadratic regression | 1% | (17%) | 7% | (26%) | -29% | (56%) | -10% | (31%) | 21% | (69%) | 1% | (13%) | -2% | (56%) |
| | Truncated regression | 7349% | (58009%) | 10623% | (75433%) | -20% | (38%) | -11% | (28%) | 27% | (75%) | -1% | (11%) | 0% | (53%) |
| | Transform X | 6% | (20%) | 29% | (57%) | 24% | (50%) | -20% | (32%) | 39% | (69%) | 16% | (22%) | -16% | (49%) |
| | Transform all | -5% | (17%) | -2% | (28%) | 0% | (28%) | -21% | (31%) | 53% | (75%) | 26% | (32%) | -33% | (59%) |
| 4 | Linear regression | -4% | (11%) | -2% | (18%) | -52% | (57%) | -2% | (24%) | 20% | (102%) | 0% | (12%) | -2% | (55%) |
| | Linear regression, censored | -2% | (10%) | -6% | (16%) | -38% | (46%) | 1% | (23%) | 3% | (99%) | 0% | (12%) | -3% | (55%) |
| | Linear regression, truncated | 0% | (11%) | -8% | (17%) | -35% | (44%) | 1% | (24%) | -6% | (100%) | 2% | (11%) | -5% | (56%) |
| | Quadratic regression | 0% | (12%) | 4% | (20%) | -31% | (57%) | -11% | (29%) | 43% | (121%) | 0% | (12%) | -3% | (56%) |
| | Truncated regression | -1% | (11%) | -3% | (16%) | -26% | (38%) | -1% | (23%) | 9% | (96%) | -1% | (11%) | -3% | (55%) |
| | Transform X | 2% | (12%) | 12% | (28%) | 23% | (50%) | -15% | (28%) | 56% | (117%) | 10% | (18%) | -11% | (51%) |
| | Transform all | -5% | (13%) | -5% | (17%) | -7% | (26%) | -16% | (28%) | 80% | (126%) | 20% | (29%) | -26% | (57%) |
| 8 | Linear regression | -1% | (7%) | 0% | (15%) | -50% | (57%) | -1% | (21%) | 23% | (180%) | 0% | (11%) | 0% | (53%) |
| | Linear regression, censored | -1% | (7%) | -2% | (14%) | -43% | (52%) | 0% | (21%) | 12% | (181%) | 0% | (11%) | 0% | (53%) |
| | Linear regression, truncated | 0% | (7%) | -3% | (13%) | -38% | (48%) | 0% | (21%) | 3% | (177%) | 1% | (10%) | -1% | (54%) |
| | Quadratic regression | 0% | (8%) | 4% | (17%) | -32% | (60%) | -9% | (26%) | 68% | (221%) | 1% | (12%) | -1% | (54%) |
| | Truncated regression | -1% | (7%) | -2% | (14%) | -37% | (47%) | 1% | (21%) | 4% | (179%) | -1% | (10%) | 1% | (53%) |
| | Transform X | 1% | (8%) | 7% | (19%) | 23% | (50%) | -9% | (24%) | 69% | (196%) | 6% | (14%) | -5% | (51%) |
| | Transform all | -4% | (9%) | -4% | (15%) | -14% | (32%) | -12% | (26%) | 125% | (219%) | 17% | (28%) | -21% | (55%) |

*Table A-1*b. By coefficient of determination

| $R^2_{Y|X}$ | Imputation method | Marginal distribution of incomplete X | | | | | | Regression of complete Y on incomplete X | | | | | | | |
|---|---|---|---|---|---|---|---|---|---|---|---|---|---|---|---|
| | | Mean | | SD | | Skew | | Slope | | Intercept | | Residual SD | | $R^2_{Y|X}$ | |
| | | Rel. bias | Rel. RMSE | Rel. bias | Rel. RMSE | Rel. bias | Rel. RMSE | Rel. bias | Rel. RMSE | Rel. bias | Rel. RMSE | Rel. bias | Rel. RMSE | Rel. bias | Rel. RMSE |
| .1 | Linear regression | -3% | (15%) | 0% | (22%) | -62% | (64%) | -5% | (52%) | 35% | (202%) | -2% | (9%) | -75% | (77%) |
| | Linear regression, censored | 2% | (15%) | -7% | (20%) | -43% | (48%) | -1% | (54%) | 18% | (213%) | -2% | (9%) | -76% | (78%) |
| | Linear regression, truncated | 10% | (20%) | -9% | (20%) | -42% | (48%) | -4% | (55%) | 10% | (218%) | -1% | (8%) | -79% | (81%) |
| | Quadratic regression | 0% | (18%) | 4% | (24%) | -62% | (66%) | -20% | (66%) | 71% | (254%) | -3% | (10%) | -72% | (75%) |
| | Truncated regression | 58350% | (557334%) | 63227% | (577610%) | -45% | (53%) | -21% | (59%) | 66% | (249%) | -3% | (10%) | -72% | (75%) |
| | Transform X | 1% | (16%) | 4% | (21%) | 1% | (23%) | -17% | (51%) | 59% | (202%) | -1% | (8%) | -79% | (80%) |
| | Transform all | -1% | (16%) | 4% | (28%) | 0% | (27%) | -27% | (53%) | 99% | (211%) | 0% | (8%) | -83% | (84%) |
| .3 | Linear regression | -7% | (16%) | -1% | (24%) | -64% | (66%) | -2% | (26%) | 23% | (106%) | 0% | (11%) | -41% | (48%) |
| | Linear regression, censored | -1% | (14%) | -8% | (20%) | -45% | (51%) | 4% | (27%) | -1% | (103%) | 1% | (10%) | -43% | (49%) |
| | Linear regression, truncated | 4% | (17%) | -10% | (22%) | -46% | (51%) | 2% | (28%) | -7% | (105%) | 2% | (11%) | -47% | (53%) |
| | Quadratic regression | 0% | (16%) | 1% | (22%) | -51% | (60%) | -13% | (37%) | 48% | (138%) | 1% | (12%) | -45% | (53%) |
| | Truncated regression | -371% | (6935%) | 1031% | (7329%) | -40% | (50%) | -14% | (35%) | 25% | (118%) | 0% | (11%) | -43% | (49%) |
| | Transform X | -1% | (14%) | 3% | (20%) | 1% | (26%) | -13% | (28%) | 41% | (107%) | 3% | (10%) | -49% | (54%) |
| | Transform all | -4% | (16%) | 3% | (31%) | -3% | (28%) | -26% | (37%) | 90% | (138%) | 7% | (13%) | -62% | (65%) |
| .5 | Linear regression | -8% | (16%) | -2% | (24%) | -59% | (62%) | 0% | (20%) | 17% | (76%) | 1% | (13%) | -5% | (29%) |
| | Linear regression, censored | -2% | (14%) | -9% | (21%) | -41% | (48%) | 6% | (20%) | -2% | (67%) | 2% | (12%) | -7% | (28%) |
| | Linear regression, truncated | 0% | (16%) | -9% | (23%) | -43% | (50%) | 5% | (21%) | -10% | (71%) | 3% | (13%) | -10% | (29%) |
| | Quadratic regression | 0% | (14%) | 2% | (22%) | -32% | (53%) | -8% | (23%) | 29% | (82%) | 2% | (14%) | -8% | (31%) |
| | Truncated regression | 1966680% | (19760437%) | 2022662% | (20322372%) | -27% | (41%) | -6% | (22%) | 5% | (66%) | 2% | (12%) | -7% | (27%) |
| | Transform X | 0% | (14%) | 6% | (25%) | 10% | (35%) | -13% | (21%) | 38% | (70%) | 6% | (13%) | -16% | (30%) |
| | Transform all | -5% | (17%) | 3% | (38%) | -2% | (31%) | -25% | (31%) | 84% | (109%) | 16% | (23%) | -40% | (50%) |
| .7 | Linear regression | -8% | (15%) | -6% | (22%) | -50% | (55%) | 7% | (21%) | 4% | (58%) | 2% | (15%) | 33% | (41%) |
| | Linear regression, censored | -4% | (13%) | -10% | (21%) | -38% | (45%) | 11% | (20%) | -11% | (51%) | 2% | (14%) | 33% | (41%) |
| | Linear regression, truncated | -4% | (15%) | -10% | (22%) | -39% | (46%) | 11% | (21%) | -15% | (56%) | 3% | (15%) | 31% | (40%) |
| | Quadratic regression | -1% | (14%) | 4% | (25%) | -16% | (52%) | -5% | (17%) | 18% | (55%) | 0% | (13%) | 35% | (42%) |
| | Truncated regression | 17% | (95%) | 43% | (236%) | -21% | (37%) | -1% | (15%) | -2% | (47%) | 1% | (13%) | 35% | (41%) |
| | Transform X | 3% | (17%) | 20% | (53%) | 26% | (53%) | -15% | (21%) | 38% | (63%) | 12% | (18%) | 20% | (31%) |
| | Transform all | -8% | (17%) | -6% | (29%) | -9% | (30%) | -17% | (23%) | 65% | (83%) | 30% | (38%) | -10% | (44%) |
| .9 | Linear regression | -3% | (12%) | -3% | (18%) | -24% | (38%) | 6% | (14%) | -2% | (34%) | 2% | (14%) | 78% | (78%) |
| | Linear regression, censored | -2% | (11%) | -5% | (17%) | -19% | (36%) | 8% | (13%) | -8% | (31%) | 2% | (14%) | 78% | (79%) |
| | Linear regression, truncated | -2% | (12%) | -5% | (17%) | -20% | (36%) | 8% | (14%) | -10% | (33%) | 2% | (14%) | 78% | (78%) |
| | Quadratic regression | 3% | (14%) | 15% | (33%) | 6% | (54%) | -8% | (13%) | 18% | (39%) | 5% | (16%) | 77% | (77%) |
| | Truncated regression | 2% | (11%) | 3% | (27%) | -11% | (33%) | 0% | (8%) | -4% | (27%) | 0% | (11%) | 79% | (79%) |
| | Transform X | 24% | (42%) | 104% | (191%) | 70% | (103%) | -26% | (31%) | 60% | (87%) | 48% | (53%) | 54% | (56%) |
| | Transform all | -7% | (15%) | -8% | (22%) | -10% | (30%) | -5% | (13%) | 31% | (47%) | 68% | (78%) | 35% | (52%) |

*Table A-1*c. By location of missing values

| Location of missing values | Imputation method | Marginal distribution of incomplete X | | | | | | Regression of complete Y on incomplete X | | | | | | | |
|---|---|---|---|---|---|---|---|---|---|---|---|---|---|---|---|
| | | Mean | | SD | | Skew | | Slope of X | | Intercept | | Residual SD | | $R^2_{Y\|X}$ | |
| | | Rel. bias | Rel. RMSE | Rel. bias | Rel. RMSE | Rel. bias | Rel. RMSE | Rel. bias | Rel. RMSE | Rel. bias | Rel. RMSE | Rel. bias | Rel. RMSE | Rel. bias | Rel. RMSE |
| MCAR | Linear regression | 0% | (12%) | -1% | (16%) | -47% | (53%) | -1% | (20%) | 4% | (77%) | 0% | (10%) | -1% | (54%) |
| | Linear regression, censored | 3% | (12%) | -4% | (16%) | -35% | (43%) | 2% | (21%) | -8% | (81%) | 0% | (10%) | -1% | (54%) |
| | Linear regression, truncated | 8% | (15%) | -6% | (16%) | -33% | (42%) | 2% | (21%) | -16% | (83%) | 1% | (10%) | -4% | (55%) |
| | Quadratic regression | 0% | (12%) | 2% | (18%) | -29% | (51%) | -4% | (21%) | 13% | (80%) | 0% | (10%) | -2% | (54%) |
| | Truncated regression | 18777% | (1871148%) | 17903% | (1747636%) | -27% | (42%) | -7% | (24%) | 13% | (90%) | -1% | (10%) | -1% | (55%) |
| | Transform X | 4% | (16%) | 19% | (46%) | 24% | (51%) | -16% | (28%) | 41% | (94%) | 15% | (21%) | -14% | (48%) |
| | Transform all | 1% | (12%) | 2% | (18%) | -5% | (26%) | -19% | (28%) | 58% | (99%) | 23% | (29%) | -28% | (54%) |
| peak | Linear regression | -10% | (17%) | 15% | (23%) | -54% | (60%) | -9% | (20%) | 54% | (95%) | -6% | (12%) | 9% | (53%) |
| | Linear regression, censored | 0% | (12%) | 4% | (16%) | -27% | (39%) | -3% | (19%) | 26% | (81%) | -4% | (10%) | 4% | (54%) |
| | Linear regression, truncated | 2% | (18%) | 4% | (18%) | -30% | (42%) | -10% | (22%) | 30% | (89%) | -2% | (11%) | 0% | (57%) |
| | Quadratic regression | 0% | (15%) | 10% | (20%) | -45% | (55%) | -20% | (30%) | 76% | (120%) | 2% | (12%) | -6% | (58%) |
| | Truncated regression | 16062% | (1515736%) | 19850% | (1725046%) | -24% | (40%) | -9% | (23%) | 41% | (102%) | -3% | (10%) | 4% | (51%) |
| | Transform X | 2% | (13%) | 3% | (16%) | -1% | (29%) | -3% | (21%) | 9% | (84%) | 4% | (13%) | -2% | (50%) |
| | Transform all | 3% | (14%) | 18% | (43%) | -1% | (38%) | -10% | (25%) | 21% | (85%) | 1% | (15%) | -4% | (55%) |
| tail | Linear regression | -8% | (16%) | -22% | (27%) | -54% | (59%) | 14% | (40%) | -12% | (114%) | 8% | (15%) | -13% | (57%) |
| | Linear regression, censored | -7% | (16%) | -23% | (27%) | -50% | (55%) | 17% | (40%) | -20% | (118%) | 7% | (15%) | -12% | (57%) |
| | Linear regression, truncated | -5% | (15%) | -24% | (28%) | -50% | (55%) | 22% | (40%) | -33% | (117%) | 7% | (15%) | -12% | (56%) |
| | Quadratic regression | 1% | (20%) | 5% | (37%) | -19% | (65%) | -9% | (43%) | 20% | (140%) | 1% | (17%) | 0% | (55%) |
| | Truncated regression | 1179968% | (11856166%) | 1214426% | (12197278%) | -35% | (47%) | -8% | (36%) | -1% | (113%) | 3% | (14%) | -7% | (57%) |
| | Transform X | 11% | (33%) | 60% | (124%) | 42% | (64%) | -31% | (42%) | 91% | (140%) | 23% | (28%) | -26% | (52%) |
| | Transform all | -19% | (22%) | -23% | (27%) | -9% | (24%) | -31% | (41%) | 143% | (168%) | 49% | (53%) | -64% | (68%) |

Table A-2. *Results of 3-variable simulation, by manipulated factors*

*Table A-2*a. By degrees of freedom in the *X* distribution

| | | Marginal distribution of incomplete X | | | | | | Regression of complete Z on incomplete X and complete Y | | | | | | | | | | | |
|---|---|---|---|---|---|---|---|---|---|---|---|---|---|---|---|---|---|---|---|
| | | Mean | | SD | | Skew | | Slope of X | | Slope of Y | | Intercept | | Residual SD | | $R^2_{Z\|XY}$ | |
| ν | Imputation method | Rel. bias | Rel. RMSE | Rel. bias | Rel. RMSE | Rel. bias | Rel. RMSE | Rel. bias | Rel. RMSE | Rel. bias | Rel. RMSE | Rel. bias | Rel. RMSE | Rel. bias | Rel. RMSE | Rel. bias | Rel. RMSE |
| 1 | Linear regression | -12% | (24%) | -6% | (30%) | -54% | (58%) | 0% | (69%) | 4% | (41%) | 1% | (72%) | -1% | (8%) | 1% | (19%) |
| | Linear regression, censored | -1% | (21%) | -15% | (28%) | -35% | (42%) | 6% | (71%) | 6% | (38%) | -19% | (67%) | 0% | (8%) | 0% | (19%) |
| | Linear regression, truncated | 3% | (27%) | -13% | (30%) | -42% | (48%) | 0% | (69%) | 8% | (40%) | -18% | (78%) | 0% | (8%) | 0% | (19%) |
| | Quadratic regression | 0% | (25%) | 9% | (36%) | -31% | (56%) | -14% | (66%) | 8% | (39%) | -1% | (66%) | -1% | (8%) | 1% | (19%) |
| | Truncated regression | 315735696% | (243045529%) | 2857099936% | (218730427896%) | -29% | (47%) | -22% | (66%) | 4% | (38%) | 18% | (77%) | -1% | (8%) | 1% | (19%) |
| | Transform X | 14% | (42%) | 66% | (140%) | 18% | (44%) | -37% | (60%) | 24% | (37%) | -11% | (59%) | 1% | (7%) | -2% | (18%) |
| | Transform all | -4% | (28%) | 8% | (64%) | 2% | (31%) | -45% | (63%) | 30% | (41%) | -13% | (61%) | 2% | (7%) | -3% | (18%) |
| 2 | Linear regression | -5% | (15%) | -2% | (23%) | -51% | (56%) | -4% | (66%) | 4% | (43%) | 2% | (103%) | -1% | (8%) | 2% | (17%) |
| | Linear regression, censored | 0% | (14%) | -8% | (20%) | -33% | (42%) | 1% | (66%) | 5% | (40%) | -18% | (98%) | -1% | (8%) | 2% | (17%) |
| | Linear regression, truncated | 3% | (17%) | -9% | (22%) | -35% | (44%) | -3% | (66%) | 8% | (41%) | -24% | (108%) | 0% | (8%) | 1% | (17%) |
| | Quadratic regression | 1% | (16%) | 8% | (26%) | -27% | (55%) | -15% | (64%) | 8% | (41%) | 3% | (98%) | -1% | (8%) | 2% | (17%) |
| | Truncated regression | 32% | (314%) | 123% | (559%) | -17% | (35%) | -8% | (62%) | 3% | (39%) | 6% | (97%) | -1% | (8%) | 2% | (17%) |
| | Transform X | 7% | (21%) | 33% | (62%) | 27% | (53%) | -34% | (58%) | 21% | (36%) | 1% | (88%) | 0% | (8%) | 0% | (16%) |
| | Transform all | -4% | (17%) | 0% | (30%) | 0% | (29%) | -36% | (59%) | 24% | (40%) | 2% | (91%) | 1% | (8%) | -2% | (17%) |
| 4 | Linear regression | -4% | (11%) | -1% | (18%) | -51% | (57%) | -4% | (64%) | 4% | (42%) | 10% | (163%) | -1% | (8%) | 1% | (16%) |
| | Linear regression, censored | -2% | (10%) | -5% | (16%) | -38% | (45%) | 1% | (65%) | 3% | (41%) | -11% | (159%) | -1% | (7%) | 1% | (16%) |
| | Linear regression, truncated | 0% | (11%) | -7% | (16%) | -34% | (44%) | 0% | (64%) | 4% | (41%) | -23% | (164%) | -1% | (8%) | 1% | (16%) |
| | Quadratic regression | 0% | (12%) | 5% | (20%) | -30% | (58%) | -12% | (64%) | 6% | (41%) | 20% | (163%) | -1% | (8%) | 1% | (17%) |
| | Truncated regression | -1% | (10%) | -3% | (15%) | -26% | (37%) | 1% | (64%) | 1% | (41%) | -5% | (155%) | -1% | (7%) | 1% | (16%) |
| | Transform X | 2% | (12%) | 13% | (29%) | 26% | (53%) | -23% | (57%) | 15% | (37%) | 15% | (147%) | 0% | (7%) | 0% | (16%) |
| | Transform all | -5% | (12%) | -4% | (18%) | -6% | (27%) | -25% | (59%) | 18% | (40%) | 22% | (153%) | 0% | (7%) | -1% | (16%) |
| 8 | Linear regression | -1% | (7%) | 0% | (15%) | -50% | (57%) | 0% | (61%) | 3% | (41%) | -15% | (292%) | -1% | (8%) | 3% | (16%) |
| | Linear regression, censored | -1% | (7%) | -1% | (14%) | -43% | (51%) | 0% | (62%) | 3% | (41%) | -24% | (297%) | -1% | (8%) | 3% | (16%) |
| | Linear regression, truncated | 0% | (7%) | -3% | (13%) | -38% | (48%) | 0% | (62%) | 4% | (41%) | -34% | (296%) | 0% | (8%) | 3% | (16%) |
| | Quadratic regression | 0% | (8%) | 6% | (18%) | -30% | (62%) | -10% | (63%) | 7% | (41%) | 14% | (293%) | -1% | (8%) | 3% | (17%) |
| | Truncated regression | -1% | (7%) | -2% | (13%) | -36% | (47%) | 3% | (63%) | 1% | (41%) | -35% | (296%) | -1% | (8%) | 3% | (16%) |
| | Transform X | 2% | (8%) | 8% | (20%) | 26% | (54%) | -14% | (55%) | 11% | (37%) | 8% | (273%) | 0% | (8%) | 2% | (16%) |
| | Transform all | -3% | (8%) | -3% | (15%) | -14% | (34%) | -20% | (59%) | 15% | (43%) | 34% | (280%) | 0% | (8%) | 1% | (16%) |

*Table A-2*b. By squared correlation between regressors

| | | Marginal distribution of incomplete X | | | | | Regression of complete Z on incomplete X and complete Y | | | | | | | | | |
|---|---|---|---|---|---|---|---|---|---|---|---|---|---|---|---|---|---|
| | | Mean | | SD | | Skew | | Slope of X | | Slope of Y | | Intercept | | Residual SD | | $R^2_{Z\|XY}$ | |
| $\rho^2_{YX}$ | Imputation method | Rel. bias | Rel. RMSE | Rel. bias | Rel. RMSE | Rel. bias | Rel. RMSE | Rel. bias | Rel. RMSE | Rel. bias | Rel. RMSE | Rel. bias | Rel. RMSE | Rel. bias | Rel. RMSE | Rel. bias | Rel. RMSE |
| .1 | Linear regression | -3% | (15%) | 1% | (22%) | -63% | (65%) | 1% | (62%) | 0% | (15%) | -12% | (220%) | -1% | (8%) | 2% | (16%) |
| | Linear regression, censored | 3% | (15%) | -6% | (19%) | -43% | (48%) | 6% | (64%) | 1% | (14%) | -33% | (231%) | -1% | (8%) | 2% | (16%) |
| | Linear regression, truncated | 10% | (19%) | -8% | (20%) | -42% | (48%) | 2% | (63%) | 2% | (14%) | -43% | (239%) | 0% | (8%) | 1% | (16%) |
| | Quadratic regression | 0% | (18%) | 7% | (24%) | -60% | (66%) | -9% | (64%) | 2% | (15%) | 12% | (226%) | -1% | (8%) | 2% | (16%) |
| | Truncated regression | 2185% | (157831%) | 241170% | (17262%) | -44% | (54%) | -14% | (65%) | 5% | (12%) | 5% | (242%) | -1% | (8%) | 2% | (16%) |
| | Transform X | 2% | (15%) | 6% | (21%) | 2% | (23%) | -11% | (56%) | 2% | (14%) | 7% | (205%) | 0% | (7%) | 1% | (16%) |
| | Transform all | 1% | (17%) | 8% | (39%) | 3% | (32%) | -26% | (56%) | 5% | (14%) | 56% | (205%) | 1% | (7%) | -1% | (16%) |
| .3 | Linear regression | -6% | (15%) | -1% | (23%) | -63% | (66%) | -1% | (47%) | 1% | (22%) | 14% | (150%) | -1% | (8%) | 2% | (17%) |
| | Linear regression, censored | -1% | (13%) | -8% | (20%) | -45% | (50%) | 2% | (49%) | 3% | (21%) | -9% | (150%) | 0% | (8%) | 1% | (17%) |
| | Linear regression, truncated | 3% | (16%) | -9% | (21%) | -45% | (50%) | -1% | (47%) | 4% | (21%) | -20% | (152%) | 0% | (8%) | 0% | (17%) |
| | Quadratic regression | 0% | (16%) | 3% | (21%) | -47% | (58%) | -10% | (50%) | 4% | (23%) | 18% | (151%) | 0% | (8%) | 1% | (17%) |
| | Truncated regression | 394646865% | (303790270496) | 357112293% | (273394904896) | -33% | (46%) | -8% | (49%) | 3% | (21%) | 3% | (150%) | -1% | (8%) | 2% | (17%) |
| | Transform X | 6% | (14%) | 6% | (21%) | 4% | (28%) | -16% | (44%) | 6% | (21%) | 22% | (136%) | 0% | (8%) | 0% | (16%) |
| | Transform all | -3% | (17%) | 5% | (39%) | -3% | (29%) | -28% | (49%) | 12% | (24%) | 47% | (138%) | 2% | (8%) | -3% | (16%) |
| .5 | Linear regression | -8% | (15%) | -2% | (23%) | -58% | (61%) | -3% | (52%) | 4% | (32%) | 3% | (144%) | -1% | (8%) | 2% | (17%) |
| | Linear regression, censored | -2% | (13%) | -9% | (20%) | -41% | (47%) | 1% | (53%) | 5% | (31%) | -20% | (139%) | 0% | (8%) | 1% | (17%) |
| | Linear regression, truncated | 0% | (16%) | -9% | (22%) | -41% | (49%) | -1% | (53%) | 7% | (32%) | -27% | (148%) | 0% | (8%) | 0% | (17%) |
| | Quadratic regression | 0% | (14%) | 4% | (22%) | -29% | (56%) | -11% | (53%) | 7% | (32%) | 4% | (134%) | -1% | (8%) | 2% | (17%) |
| | Truncated regression | 927% | (8907%) | 1075% | (9221%) | -25% | (39%) | -5% | (52%) | 4% | (30%) | -13% | (137%) | -1% | (8%) | 2% | (17%) |
| | Transform X | 1% | (14%) | 9% | (28%) | 14% | (40%) | -22% | (49%) | 13% | (31%) | 8% | (125%) | 0% | (7%) | 0% | (17%) |
| | Transform all | -5% | (16%) | 0% | (29%) | -4% | (31%) | -32% | (53%) | 22% | (36%) | 10% | (126%) | 2% | (8%) | -3% | (17%) |
| .7 | Linear regression | -8% | (15%) | -6% | (21%) | -49% | (54%) | -2% | (66%) | 5% | (49%) | 8% | (134%) | -1% | (7%) | 1% | (18%) |
| | Linear regression, censored | -4% | (13%) | -10% | (20%) | -37% | (44%) | 3% | (65%) | 5% | (46%) | -6% | (126%) | -1% | (7%) | 0% | (18%) |
| | Linear regression, truncated | -3% | (15%) | -10% | (21%) | -38% | (46%) | 0% | (65%) | 7% | (47%) | -10% | (134%) | -1% | (7%) | 0% | (18%) |
| | Quadratic regression | -1% | (14%) | 5% | (25%) | -17% | (54%) | -13% | (62%) | 7% | (47%) | 19% | (126%) | -1% | (8%) | 1% | (18%) |
| | Truncated regression | 8% | (72%) | 28% | (166%) | -21% | (36%) | -3% | (61%) | 2% | (45%) | 1% | (122%) | -1% | (7%) | 1% | (18%) |
| | Transform X | 4% | (19%) | 25% | (60%) | 31% | (59%) | -30% | (58%) | 19% | (44%) | 13% | (115%) | 0% | (7%) | -1% | (17%) |
| | Transform all | -7% | (17%) | -5% | (30%) | -9% | (29%) | -32% | (62%) | 29% | (51%) | -6% | (124%) | 0% | (7%) | -2% | (18%) |
| .9 | Linear regression | -3% | (11%) | -3% | (17%) | -24% | (38%) | -5% | (99%) | 9% | (90%) | -16% | (139%) | -1% | (8%) | 3% | (18%) |
| | Linear regression, censored | -1% | (11%) | -5% | (17%) | -19% | (36%) | -2% | (97%) | 8% | (86%) | -22% | (131%) | -1% | (7%) | 3% | (18%) |
| | Linear regression, truncated | -2% | (12%) | -5% | (17%) | -20% | (36%) | -4% | (98%) | 9% | (88%) | -24% | (135%) | -1% | (8%) | 3% | (18%) |
| | Quadratic regression | 3% | (14%) | 15% | (32%) | 6% | (53%) | -21% | (93%) | 16% | (87%) | -8% | (139%) | -1% | (8%) | 3% | (18%) |
| | Truncated regression | 2% | (11%) | 2% | (20%) | -11% | (33%) | -1% | (94%) | 2% | (87%) | -14% | (133%) | -1% | (7%) | 3% | (18%) |
| | Transform X | 24% | (41%) | 106% | (183%) | 71% | (104%) | -56% | (81%) | 48% | (74%) | -35% | (128%) | 0% | (7%) | 1% | (18%) |
| | Transform all | -6% | (20%) | -6% | (22%) | -8% | (30%) | -39% | (81%) | 41% | (79%) | -50% | (139%) | 2% | (8%) | 2% | (18%) |

*Table A-2*c. By location of missing values

| | | Marginal distribution of incomplete X | | | | | | Regression of complete Z on incomplete X and complete Y | | | | | | | | | |
|---|---|---|---|---|---|---|---|---|---|---|---|---|---|---|---|---|---|
| | | Mean | | SD | | Skew | | Slope of X | | Slope of Y | | Intercept | | Residual SD | | $R^2_{Z\|XY}$ | |
| Location of missing values | Imputation method | Rel. bias | Rel. RMSE | Rel. bias | Rel. RMSE | Rel. bias | Rel. RMSE | Rel. bias | Rel. RMSE | Rel. bias | Rel. RMSE | Rel. bias | Rel. RMSE | Rel. bias | Rel. RMSE | Rel. bias | Rel. RMSE |
| MCAR | Linear regression | 0% | (12%) | 0% | (16%) | -47% | (53%) | -4% | (62%) | 3% | (40%) | -5% | (149%) | -1% | (8%) | 2% | (17%) |
| | Linear regression, censored | 3% | (12%) | -4% | (16%) | -34% | (43%) | 0% | (63%) | 3% | (39%) | -21% | (152%) | -1% | (8%) | 2% | (17%) |
| | Linear regression, truncated | 7% | (14%) | -6% | (16%) | -32% | (42%) | -3% | (62%) | 5% | (39%) | -29% | (156%) | 0% | (8%) | 1% | (17%) |
| | Quadratic regression | 1% | (11%) | 3% | (18%) | -27% | (51%) | -9% | (60%) | 6% | (38%) | 1% | (148%) | -1% | (8%) | 2% | (17%) |
| | Truncated regression | 57% | (467%) | 187% | (814%) | -25% | (40%) | -7% | (62%) | 2% | (39%) | -3% | (150%) | -1% | (8%) | 2% | (17%) |
| | Transform X | 4% | (16%) | 21% | (49%) | 27% | (54%) | -30% | (56%) | 19% | (36%) | 3% | (139%) | 0% | (7%) | 0% | (17%) |
| | Transform all | 2% | (13%) | 4% | (20%) | -3% | (27%) | -34% | (56%) | 23% | (37%) | 7% | (143%) | 1% | (7%) | -1% | (17%) |
| peak | Linear regression | -9% | (16%) | 14% | (22%) | -54% | (59%) | -3% | (57%) | -6% | (44%) | 46% | (159%) | -2% | (8%) | 3% | (17%) |
| | Linear regression, censored | 1% | (12%) | 4% | (16%) | -27% | (39%) | 4% | (57%) | -4% | (40%) | 12% | (147%) | -1% | (8%) | 3% | (17%) |
| | Linear regression, truncated | 2% | (17%) | 4% | (17%) | -31% | (42%) | -1% | (57%) | -1% | (42%) | 2% | (158%) | -1% | (8%) | 2% | (17%) |
| | Quadratic regression | -1% | (15%) | 11% | (20%) | -46% | (57%) | -17% | (59%) | 6% | (41%) | 31% | (154%) | -1% | (8%) | 1% | (17%) |
| | Truncated regression | 11541% | (79181%) | 12710% | (87264%) | -23% | (39%) | -1% | (57%) | -2% | (38%) | 12% | (151%) | -1% | (8%) | 2% | (17%) |
| | Transform X | 2% | (12%) | 3% | (16%) | -2% | (29%) | -8% | (52%) | 7% | (35%) | -14% | (142%) | -1% | (7%) | 2% | (17%) |
| | Transform all | 3% | (15%) | 18% | (49%) | -2% | (39%) | -11% | (59%) | 3% | (41%) | 16% | (144%) | 0% | (8%) | 1% | (17%) |
| tail | Linear regression | -8% | (16%) | -21% | (26%) | -6% | (54%) | 7% | (76%) | 14% | (41%) | -43% | (165%) | 0% | (8%) | 2% | (17%) |
| | Linear regression, censored | -7% | (15%) | -22% | (27%) | -49% | (54%) | 2% | (77%) | 13% | (41%) | -46% | (167%) | 0% | (8%) | 0% | (17%) |
| | Linear regression, truncated | -5% | (15%) | -23% | (27%) | -49% | (54%) | 1% | (77%) | 14% | (41%) | -47% | (170%) | 0% | (8%) | 0% | (17%) |
| | Quadratic regression | 1% | (20%) | 7% | (36%) | -16% | (65%) | -12% | (74%) | 10% | (43%) | -5% | (164%) | -1% | (8%) | 2% | (18%) |
| | Truncated regression | 2367901977% | (18227620689%) | 2142696644% | (16403905719%) | -33% | (45%) | -11% | (73%) | 7% | (42%) | -19% | (166%) | -1% | (8%) | 2% | (17%) |
| | Transform X | 12% | (33%) | 66% | (124%) | 48% | (70%) | -43% | (64%) | 27% | (39%) | 20% | (144%) | 1% | (8%) | -1% | (17%) |
| | Transform all | -17% | (21%) | -21% | (26%) | -8% | (24%) | -49% | (65%) | 39% | (45%) | 11% | (152%) | 2% | (8%) | -4% | (17%) |